\ifdefined\ACMReview
  \documentclass[sigconf,anonymous,review]{acmart}
\else
  \documentclass[sigconf,nonacm]{acmart}
\fi
\makeatletter
\@ACM@balancefalse
\makeatother

\usepackage{booktabs}
\usepackage{graphicx}
\usepackage{multirow}
\usepackage{tabularx}
\usepackage{placeins}
\graphicspath{{figs/}}

\newcommand{\MW}{\ensuremath{\mathrm{MW}}}
\newcommand{\MWh}{\ensuremath{\mathrm{MWh}}}
\newcommand{\Kah}{K_{\alpha,h}}

\setcopyright{none}
\ifdefined\ACMReview
\else
\fi
\renewcommand\footnotetextcopyrightpermission[1]{}
\acmConference[e-Energy '27]{The 18th ACM International Conference on Future and Sustainable Energy Systems}{June 5--11, 2027}{Atlanta, GA, USA}
\acmBooktitle{The 18th ACM International Conference on Future and Sustainable Energy Systems (e-Energy '27), June 5--11, 2027, Atlanta, GA, USA}
\acmDOI{}
\acmISBN{}

\begin{document}

\title{Beyond Scalar Flexibility: From Eligible AI Workloads to Dependable Load Relief}

\author{Meiyi Li}
\email{Meiyi.Li@lsu.edu}
\affiliation{%
  \institution{Louisiana State University}
  \city{Baton Rouge}
  \state{Louisiana}
  \country{USA}}
\renewcommand{\shortauthors}{Li}

\begin{abstract}
Grid studies often represent data-center flexibility as a fixed percentage of load, although no public production trace has shown how much eligible load persists across event durations or co-moves across clusters. We reconstruct 4,439 hourly power observations from a 185-day trace of 155,410 GPUs and derive a workload-semantic flexibility envelope. The fleet's time-averaged Monte Carlo median facility demand is 55.8~\MW{}, while immediate eligible curtailment averages 3.55~\MW{} after retaining allocated-GPU idle power: 12.1\% of workload power and 6.35\% of median facility power. Under full realization of that eligibility, 95\%-available relief falls from 2.51~\MW{} for one hour to 2.32~\MW{} for four hours and 1.95~\MW{} for 24 hours; a common realizable fraction $q$ scales every value exactly by $q$. A mean-calibrated scalar overstates these quantities by 17\%, 25\%, and 47\%, while a scalar tail-calibrated at four hours understates the one-hour product by 6\% and overstates the 24-hour product by 17\%; the share that reproduces the surface varies by a factor of 1.6 across durations and reliability levels. Aggregating 13 clusters raises four-hour firmness from 0.38 to 0.66, but cross-cluster covariance limits the gain. The production scheduler exposes almost no additional delay-based capacity: newly deferrable arrivals average 0.008~\MW{} and have zero 95\%-available capacity. These results replace an assumed flexibility percentage with duration, reliability, portfolio, and realizability terms that can be written into interconnection and demand-response contracts.
 \end{abstract}

\ccsdesc[500]{Hardware~Power and energy}
\ccsdesc[300]{Computer systems organization~Distributed architectures}
\ccsdesc[300]{Mathematics of computing~Time series analysis}
\keywords{AI data centers, load flexibility, demand response, large-load interconnection, GPU cluster trace}

\maketitle

\section{Introduction}\label{sec:intro}

AI data-center projects are arriving faster than power systems can connect them. US data centers consumed an estimated 176~TWh in 2023, or 4.4\% of national electricity, and projections place them at 6.7--12\% by 2028~\cite{shehabi2024}. A hyperscale campus can be site-ready in roughly two years, while the substations and transmission reinforcements it needs can take five to ten~\cite{kim2026siting}. Grid institutions have therefore begun to link large-load interconnection and emergency operation to managed load reduction. Texas Senate Bill~6 establishes large-load reliability requirements, while PJM and MISO have proposed distinct connect-and-manage and flexible-service paths~\cite{ercot_sb6,pjm_cam2026,miso_flits2026}. Their central engineering question is not whether a data center contains flexible computation, but how many megawatts it can release for a stated duration and reliability.

Planning models answer that question with assumed percentages rather than production evidence. Recent capacity-expansion and siting studies divide load among fixed, shiftable, and interruptible tiers, sweep the share that can pause or move, or test whole-site curtailment against a small annual hour budget~\cite{khanal2026shift,chen2026defer,norris2025rethinking,kim2026siting}. Those abstractions have enabled system-level studies, but a scalar cannot express whether the same work remains eligible for one or 24 consecutive hours, whether eligibility co-moves across clusters, or whether the scheduler has already spent the apparent flexibility. Controlled studies establish that response is technically possible: a 256-GPU cluster shed roughly one quarter of its power for three hours, while a small-scale profiling study extrapolated 18--55\% regulation reserve through a 1,000-server simulation~\cite{colangelo2025nature,acun2026flex}. They do not measure the prevalence or persistence of eligible work in a production fleet.

Alibaba's 2026 production trace exposes the missing workload structure at hyperscale. The trace follows 155,410 GPUs on 37,707 servers in 17 clusters for 185 days, with hourly GPU allocation and utilization, public workload type, priority, and standby state, and with scheduling delay for the 47\% of execution spans that join to its execution summary~\cite{asi_trace_2026}. Earlier grid-facing studies used an older single-cluster trace, an HPC trace, inference request logs, or private normalized profiles~\cite{caprara2026ladflex,majumder2026composition,wilkins2026servers,hall2024vcc}. The new trace spans training, online inference, offline inference, and development across a hierarchy large enough to measure portfolio effects. It contains no power meter, so converting it to electrical demand requires an explicit model and an uncertainty audit rather than a claim of direct measurement.

We turn the trace into a planner-facing load and flexibility data product. Workload-conditioned GPU curves reconstruct hourly workload, IT, and facility power while 300 model draws expose parameter sensitivity and 117 compiled measurements, 92 of them in represented categories, check the anchors. A four-layer semantic envelope separates attributed workload power from the immediate active power removed when a preempted GPU remains allocated and draws its idle floor. The duration--reliability--portfolio surface $\Kah(S)$ preserves time and cluster structure when we test the fixed-share, time-shuffled, and independent-cluster representations used in planning. The result is a reproducible translation from scheduler records to contract variables, together with the assumptions needed to interpret each megawatt.

The production evidence rejects a single flexibility percentage without claiming a completed demand-response product. Immediate eligible curtailment averages 3.55~\MW{}, or 12.1\% of workload power but only 6.35\% of median facility power. If every eligible watt is realizable, 95\%-available relief falls from 2.51~\MW{} for one hour to 1.95~\MW{} for 24 hours; if only a fraction $q$ is realizable, the entire surface scales exactly by $q$. A mean-calibrated scalar overstates the one-, four-, and 24-hour products by 17\%, 25\%, and 47\%; tail calibration at four hours instead makes the one-hour product 6\% low and the 24-hour product 17\% high. Cross-cluster aggregation raises four-hour firmness from 0.38 to 0.66, although positive residual covariance prevents the independence gain. Observed queueing adds essentially no dependable capacity, and prospective rolling calibration requires a much larger derating at 24 hours than at four. Flexibility contracts should therefore specify an eligible-power boundary, duration, reliability, portfolio, response realization, recovery, and recalibration rule rather than one percentage of site load.
 \section{Trace, Power Reconstruction, and Metrics}\label{sec:methods}

\subsection{Production trace}

The public trace resolves workload semantics at an hourly scale but not electrical power. Its pod table reports allocated and used GPU-hours, mean streaming-multiprocessor utilization, CPU use, job type, priority, accelerator type, and public state; its server table reports accelerator count and model; and its execution summary reports submission, start, and end times~\cite{asi_trace_2026}. The observation window contains 4,439 valid hours over relative days 0--184 because day~5 hour~06 is absent. Seventeen clusters appear, 16 carry nonzero reconstructed GPU power, 14 meet the support rule for fluctuation scaling, and 13 carry enough eligible low-priority load for the dependable-relief portfolio. The public state is ``Unknown'' for almost every active GPU pod, so used GPU-hours rather than state define activity. Relative days prevent calendar or weather alignment.

The trace's execution join supports a narrower scheduling analysis than its power analysis. We extracted 77.83~million execution spans and matched 36.69~million, or 47.1\%, to delay records; the matched set covers 58.8\% of GPU-hours over the full trace. Match coverage rises abruptly near day~108, at the same time as the apparent change in the delay-derived difference process. We therefore discard the pre-jump period for scheduler-effect estimates and use days 110--184. A record-level audit of that restricted window matches all but 108.58 of 251.88~million GPU-hours, or 99.99996\%. This restriction removes a data-coverage artifact rather than identifying an operator policy change.

\subsection{Workload-conditioned power reconstruction}

The reconstruction separates workload, IT, and facility power. For an aggregate row with $w$ allocated GPU-hours, accelerator rating $T$, utilization $u$, idle fraction $\phi$, and workload class $k$, GPU energy follows
\begin{equation}
E_{\mathrm{GPU}}=Tw\left[\phi+(1-\phi)g_k(u)\right].
\label{eq:power}
\end{equation}
The piecewise-linear curves $g_k$ use knots at $u=\{0,0.25,0.5,0.75,1\}$ and separate training, online inference, offline inference, and other work. Online inference receives a 0.43-$T$ total-power floor at the central setting because memory-bound decoding draws substantial power at low reported SM utilization; the implementation converts this total-power anchor to the active span above idle, then applies an affine compression that preserves saturation. Standby rows draw only $\phi T$. Host power is linear in used CPU cores between 1.5 and 4.0~W per core. IT power adds host and unallocated-GPU idle power, while facility power multiplies IT power by a 1.2 central PUE. Appendix~\ref{app:power} reports all anchors.

The aggregation stores sufficient statistics rather than individual pods. For each hour, cluster, accelerator, workload type, priority, and state, we retain $S_b=\sum_p w_p\max(u_p-b,0)$ at the curve knots. Any piecewise-linear curve on those knots can then be evaluated from 858,816 aggregate rows without rescanning the 351~GB pod table. This representation makes every power-model sensitivity use the same underlying workload counts.

The uncertainty analysis treats power parameters as systematic errors. We recompute the full pipeline for 300 draws of NVIDIA rated power, unknown-accelerator rating, $\phi$, online floor, null-utilization power, host coefficients, PUE, and curve family. Draws apply to the entire series, so their P5--P95 range does not shrink as if hourly errors were independent. We also re-anchor the curves to measured H100 levels and compare central anchors with 117 public measurements, 92 of which match a represented category. The base training curve reaches rated power and can exceed measured mean training power by 15--45\%; re-anchoring and a concavity bound in Appendix~\ref{app:power} expose that uncertainty rather than hiding it.

\subsection{Workload-semantic flexibility envelope}

The envelope describes eligibility before it claims controllability. Its precedence is standby, eligible-attributed, shift, then floor. The central mapping assigns every low-priority row to the eligible-attributed layer, high-priority training and offline inference to the shift layer, Standby rows to standby, and all remaining power to the floor. A conservative mapping limits eligibility to low-priority training and offline inference; a liberal mapping also admits high-priority offline inference. The shift layer is a semantic pool, not an offered capacity, because a running job may require checkpoint, migration, or restart.

The immediate curtailment boundary retains allocated-GPU idle power. If a low-priority pod is preempted and its GPU remains powered, the removable quantity is the active term $(1-\phi)Twg_k(u)$ in Eq.~\ref{eq:power}; we call this the \emph{idle-retained} boundary. Removing idle power as well yields the attributed upper bound and requires node sleep or reassignment. Reported percentages use workload power as their denominator unless facility power is named. A marginal facility response counts one removed workload megawatt as one facility megawatt, whereas average-PUE accounting multiplies it by 1.2. These conventions prevent a workload share from masquerading as a metered-facility share; Table~\ref{tab:boundary} collects the accounting.

\begin{table}[t]
\caption{Boundary accounting at central parameters (Monte Carlo median in parentheses where it differs). Eligible-layer quantities use the central mapping.}
\label{tab:boundary}
\centering
\small
\begin{tabular}{@{}lr@{}}
\toprule
Quantity & MW \\
\midrule
GPU-side workload power (pods) & 29.37 (28.92) \\
IT power (workload + host + unallocated idle) & 43.60 (45.08) \\
Facility power (IT $\times$ PUE 1.2) & 52.32 (55.83) \\
Eligible (low-priority) attributed pod power, 15 clusters & 5.99 \\
\quad of which idle draw retained after reclaim & 2.44 \\
\quad immediate idle-retained relief, 15 clusters & 3.55 \\
\quad same, 13 clusters above the 0.02~MW threshold & 3.53 \\
Direct host relief from evicted pods' CPU (not in headline) & 0.15 \\
Facility-side relief, marginal 1.0 / average PUE 1.2 & 3.55 / 4.26 \\
\bottomrule
\end{tabular}
\end{table}

\subsection{Dependable eligible-workload availability}

The duration--reliability surface measures persistence in the observed eligibility series. Let $C_{c,t}$ denote idle-retained eligible power in cluster $c$ at hour $t$, $S$ a cluster portfolio, $h$ an event duration, and $q\in[0,1]$ the fraction of eligible power that a control system can realize. We define
\begin{equation}
K_{\alpha,h}(S;q)=\sup\left\{k:\Pr_t\!\left[\min_{0\leq j<h}q\sum_{c\in S}C_{c,t+j}\geq k\right]\geq\alpha\right\}.
\label{eq:k}
\end{equation}
We evaluate $h\in\{1,2,4,8,24\}$ hours and $\alpha\in\{0.5,0.9,0.95,0.99\}$. $K$ is an unconditional lower-tail statistic over uniformly sampled historical start hours of hour-average eligible power on the untouched baseline: it assumes that all contemporaneous low-priority work can be suppressed for $h$ hours with no refill of reclaimed GPUs, that the eligible cohort need not be the same throughout the window, and that curtailment does not alter the subsequent trace. It is not availability conditional on grid stress and not an intervention result. We report firmness as $K_{\alpha,h}(S;1)$ divided by the mean eligible power of portfolio $S$. Equation~\ref{eq:k} is positively homogeneous, so scaling eligibility by $q$ scales every $K$ by exactly $q$; the trace identifies the shape at $q=1$ but does not identify $q$ itself.

Three counterfactual representations isolate what a scalar discards. A mean-calibrated scalar multiplies hourly workload power by its mean eligible share. A time-shuffled series preserves the marginal distribution but destroys persistence. An independent-cluster series circularly shifts each cluster by an independent random offset, preserving its marginal distribution and within-cluster persistence while removing contemporaneous portfolio covariance. We also test a tail-calibrated scalar whose constant workload share is chosen so its own $K_{0.95,4}$ equals the observed four-hour product by construction. Power-model draws are paired across observed and counterfactual series, and a 168-hour moving-block bootstrap with 200 replicates, blocks resampled jointly across clusters, indicates sampling uncertainty; with 200 replicates the P5--P95 limits are themselves imprecise, and the bands do not separate parameter from temporal uncertainty. We exclude three clusters whose mean eligible power is below 0.02~\MW{} from the portfolio surface.

A rolling-origin backtest replaces the favorable half split. Each of 15 origins estimates $K$ from eight training weeks and tests it on the next four weeks. We report absolute-MW calibration in the main text and normalized-by-mean calibration as a sensitivity. A multiplicative derating is chosen so mean out-of-sample coverage meets the target; this rule produces 2.29~\MW{} at four hours and 1.73~\MW{} at 24 hours for the 95\% product under the absolute-MW specification. Because the factor is chosen on the same folds it is evaluated on, the result is retrospective calibration rather than a prospective guarantee.

\subsection{Aggregation and scheduler diagnostics}

Fluctuation scaling and synchrony quantify where portfolio assumptions fail. We subtract a centered 169-hour rolling mean from each GPU-power series and fit $\sigma\propto\mu^\beta$ across cluster--type--priority, cluster--type, and cluster units; the $\beta=1/2$ independence benchmark assumes comparable units, and the portfolio surface is the direct evidence on aggregation. For a parent with child series $i$, the synchrony ratio is $R=\sigma(\sum_iP_i)/(\sum_i\sigma_i^2)^{1/2}$; $R=1$ is the zero-covariance benchmark. Covariance attribution uses a common set of ten clusters whose weekly mean stays above both 20\% of its median and 0.05~\MW{} for at least 80\% of hours. We compare raw residuals, hour-of-day and weekday residuals, capacity-normalized residuals, and six stable-capacity plateaus.

The scheduler diagnostic separates shifted execution from newly deferrable arrivals. For each matched execution, we compare observed power with a rectangle of equal duration and power moved back to the job's arrival time, ignoring capacity and dependency constraints; their difference $D(t)$ and integral are a temporal-translation diagnostic of the net displacement already exercised, not a feasible alternative schedule, and positive and negative displacements can cancel within an hour. A stricter backlog replay counts only new high-priority training and offline-inference arrivals whose observed queueing delay reached the requested horizon. Because recorded delay reflects the policy that ran, we call it observed queueing delay rather than workload tolerance.
 \section{Results}\label{sec:results}

\subsection{A growing but persistently loaded AI fleet}\label{sec:load}

\begin{figure*}[t]
  \centering
  \includegraphics[width=0.98\textwidth]{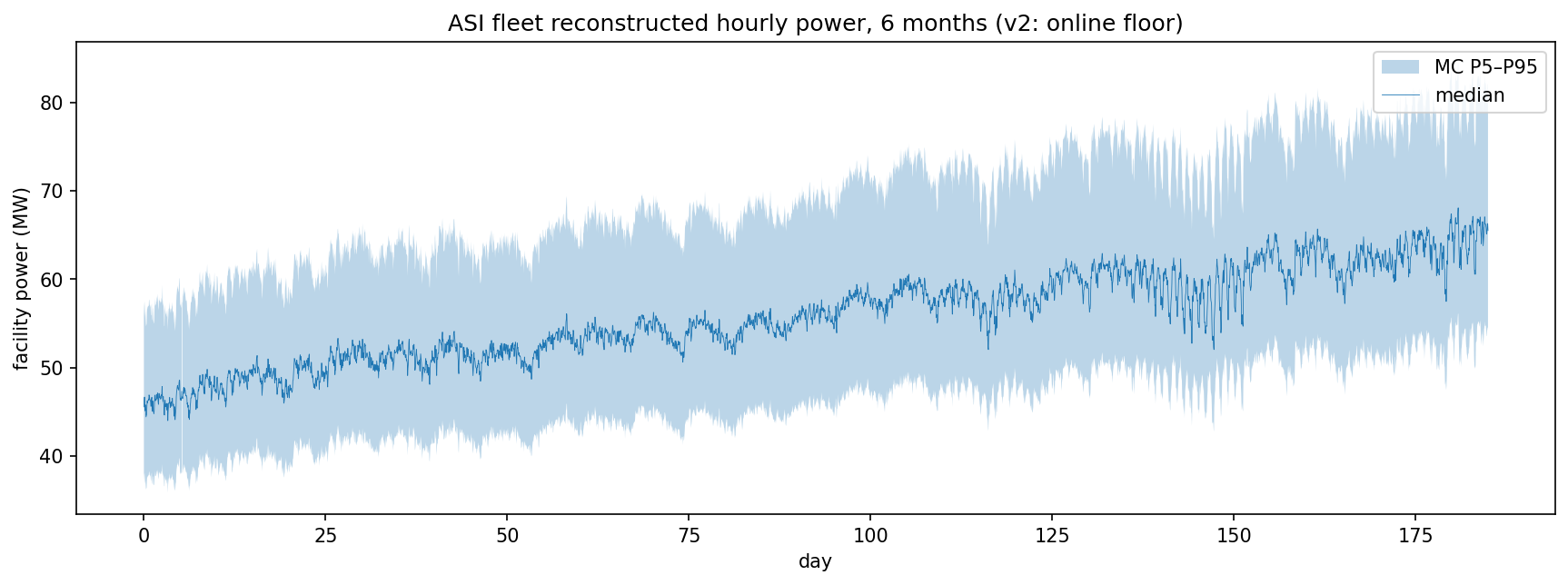}\\[2pt]
  \includegraphics[width=0.62\textwidth]{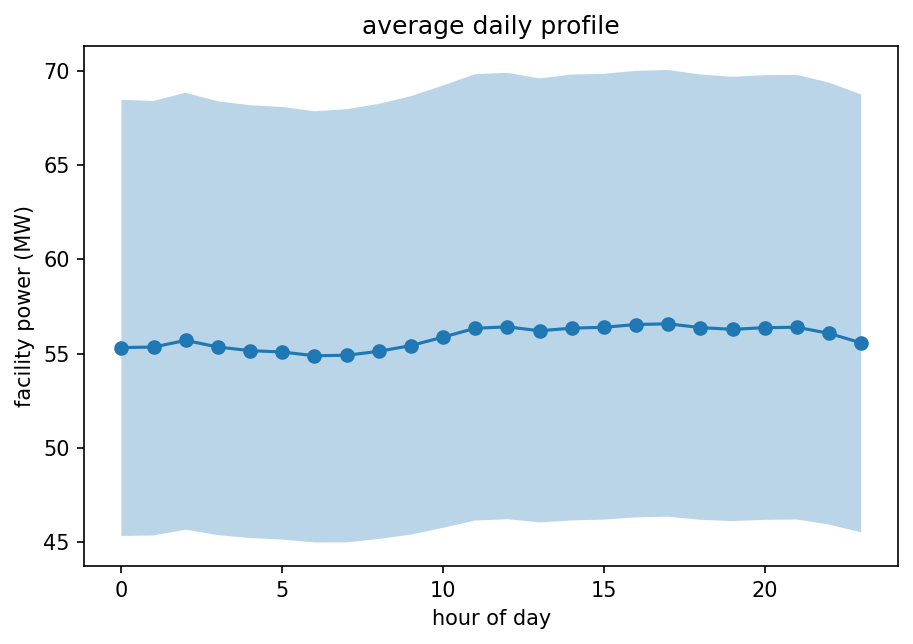}
  \caption{Reconstructed facility demand of the 155,410-GPU fleet. Top: hourly Monte Carlo median and P5--P95 power-model band over 185 days; the one missing trace hour is left blank. Bottom: clock-hour profile in trace-local time. Growth from the first to last week is 38.84\%; the mean profile troughs at hour~06 and peaks at hour~17.}
  \Description{A time series shows facility power rising in steps over 185 days with an uncertainty band. A smaller plot shows the average profile by hour of day, with a morning trough and late-afternoon peak.}
  \label{fig:load}
\end{figure*}

The reconstructed facility demand is large, flat relative to its level, and rapidly growing. Across 4,439 hours, the hourly Monte Carlo median facility series averages 55.83~\MW{}, and the time-averaged pointwise P5--P95 limits are 45.74--69.10~\MW{}; the central-parameter series averages 52.32~\MW{} (Fig.~\ref{fig:load}). Its first and last weeks average 46.28 and 64.25~\MW{}, a 38.84\% increase delivered in capacity steps. The median series has a 0.820 load factor (mean over maximum). Allocated GPU-pod power accounts for 67.35\% of central IT---and therefore central facility---power, hosts for 31.07\%, and the unallocated-GPU idle floor for 1.58\%. Near-full allocation means that low reported GPU utilization does not translate into an equally large pool of powered-off hardware.

The hourly load carries repeatable clock-time structure without resembling an inference-request trace. The peak-to-trough swing within each day averages 3.42~\MW{}, or 6.13\% of the Monte Carlo median facility power, while the profile formed by averaging each clock hour across the trace spans 3.04\%. The mean clock-hour profile troughs at hour~06 and peaks at hour~17 in trace-local time. Raw autocorrelation is 0.950 at 24 hours and 0.846 at seven days, but these values also contain the fleet's growth trend; we treat them as descriptive persistence rather than a decomposition of daily and weekly causes. Online-inference floor power alone accounts for 13.79\% of IT power, which explains why demand variation is much smaller than variation in request arrivals.

\subsection{Aggregation smooths less than independence predicts}\label{sec:aggregation}

\begin{figure*}[t]
  \centering
  \begin{minipage}[t]{0.49\textwidth}
    \centering\includegraphics[width=\linewidth]{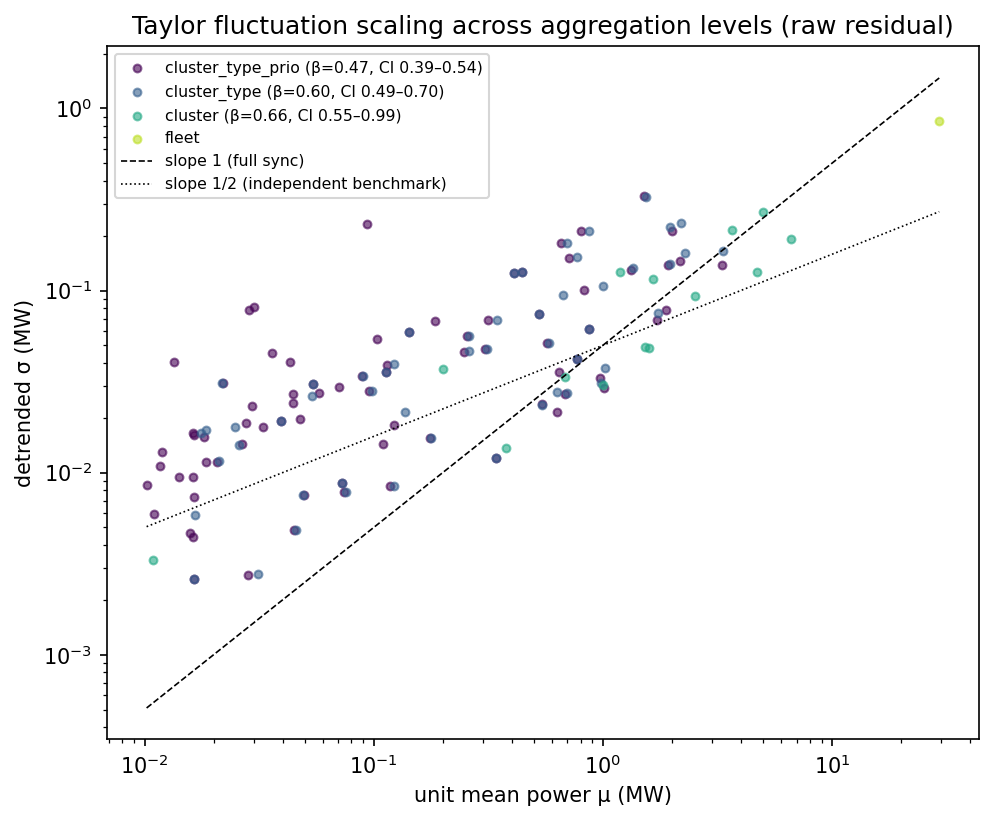}
  \end{minipage}\hfill
  \begin{minipage}[t]{0.49\textwidth}
    \centering\includegraphics[width=\linewidth]{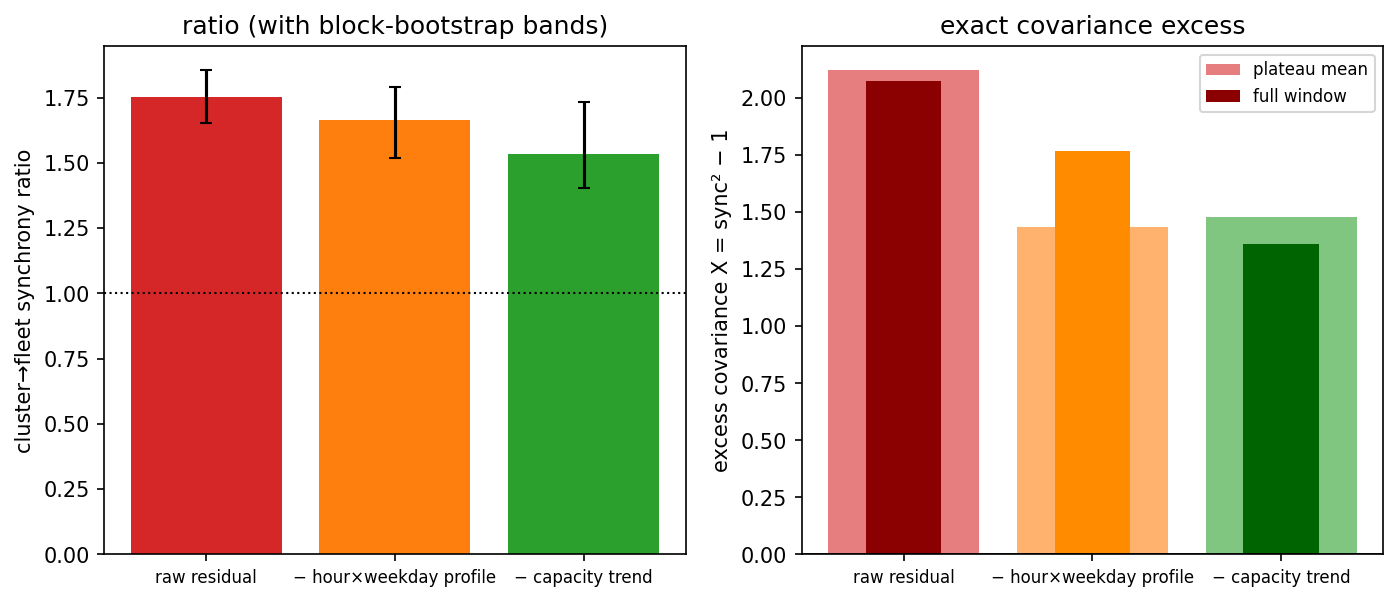}
  \end{minipage}
  \caption{Aggregation and cross-cluster covariance. (a) Fluctuation scaling across three nested unit definitions; the fitted exponent (labelled $\alpha$ in the panel, $\beta$ in the text) rises from 0.470 to 0.662. (b) Cross-cluster excess covariance on the common ten-cluster set. Calendar removal accounts for 14.6\% and capacity normalization for 34.4\% cumulatively, leaving 65.6\% unexplained.}
  \Description{The left panel is a log-log fluctuation-scaling plot at three aggregation levels. The right panel decomposes cross-cluster excess covariance into portions removed by calendar and capacity adjustments and a larger unexplained remainder.}
  \label{fig:aggregation}
\end{figure*}

Variability decreases with scale, but not at the independent-unit rate. The fluctuation-scaling exponent $\beta$ rises from 0.470 across cluster--type--priority units to 0.597 across cluster--type units and 0.662 across clusters (Fig.~\ref{fig:aggregation}a). The cluster-level 90\% bootstrap interval is 0.547--0.987, and a linear power curve gives 0.491, 0.624, and 0.666; the three intervals overlap, so the rise is a point-estimate pattern rather than an established difference. Values below one mean aggregation still smooths relative fluctuations; values above the independent benchmark of one half mean that this smoothing weakens at coarser levels.

The sign of covariance reverses across the hierarchy. Priority classes within a cluster--type group have a median synchrony ratio of 0.984, close to the zero-covariance value of one. Workload types within a cluster have a ratio of 0.654, consistent with a capacity-constrained scheduler allocating GPUs among competing classes. Clusters within the fleet have a ratio of 1.758 (jackknife 90\% interval 1.492--2.024), so their residuals reinforce one another. In a common ten-cluster subset used for guarded attribution, the ratio is 1.753 (block-bootstrap interval 1.653--1.854), and fleet residual variance is 3.07 times the sum of individual variances.

Observed calendars and capacity changes explain only one third of the cross-cluster excess covariance. Removing hour-of-day and weekday means removes 14.6\% of the excess-variance term $R^2-1$; subsequent capacity normalization raises the cumulative share removed to 34.4\% (Fig.~\ref{fig:aggregation}b). The raw ratio averaged across six stable-capacity plateaus is 1.765, almost identical to the full-window value, so synchronized expansion is not the main cause. Roughly two thirds of the excess remains unidentified. The production trace establishes a portfolio penalty relative to independence, but it does not identify a single mechanism for that penalty.

\subsection{Eligible workload is smaller than its semantic layer}\label{sec:envelope}

\begin{figure*}[t]
  \centering
  \begin{minipage}[t]{0.58\textwidth}
    \centering\includegraphics[width=\linewidth]{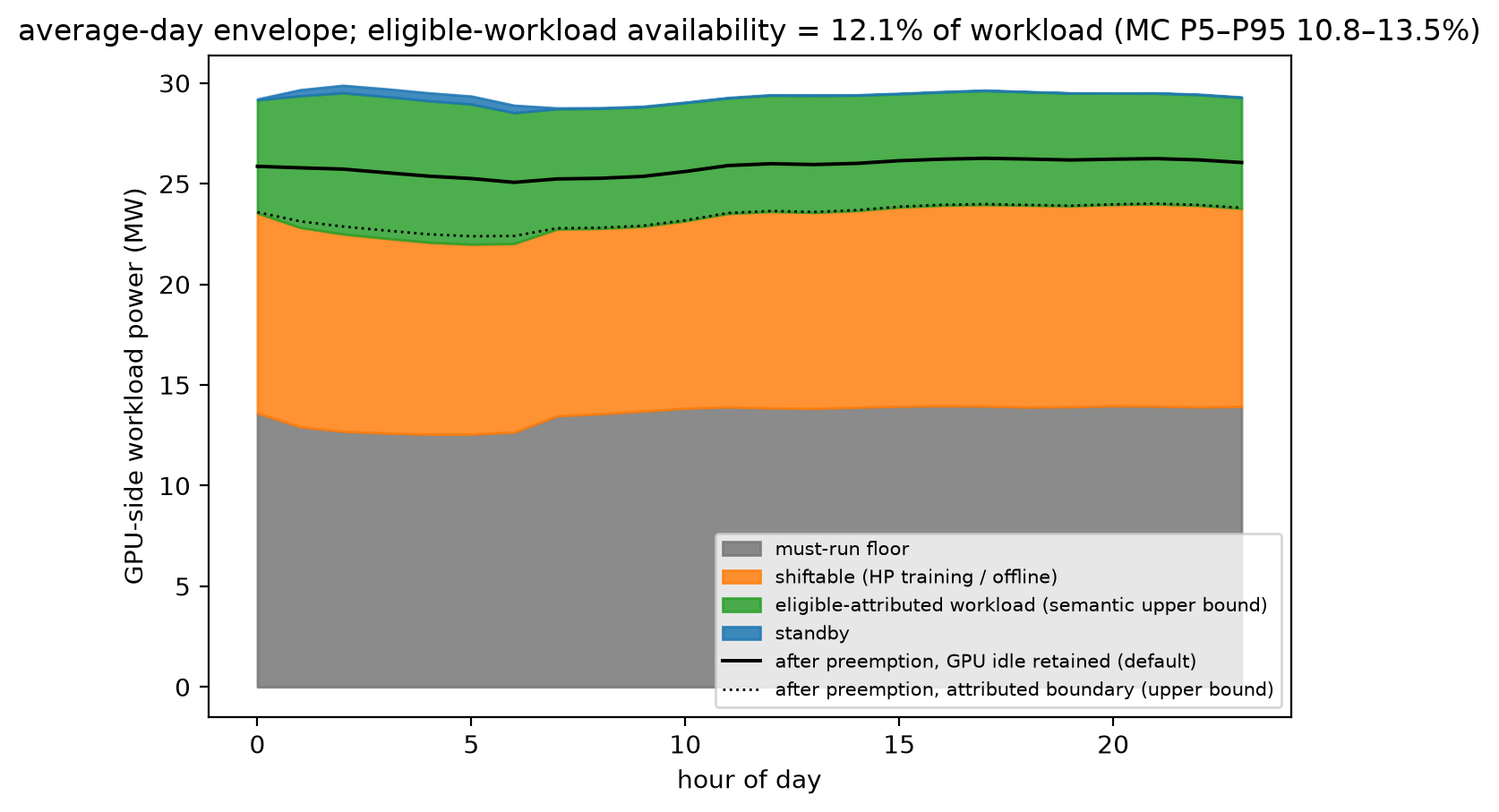}
  \end{minipage}\hfill
  \begin{minipage}[t]{0.40\textwidth}
    \centering\includegraphics[width=\linewidth]{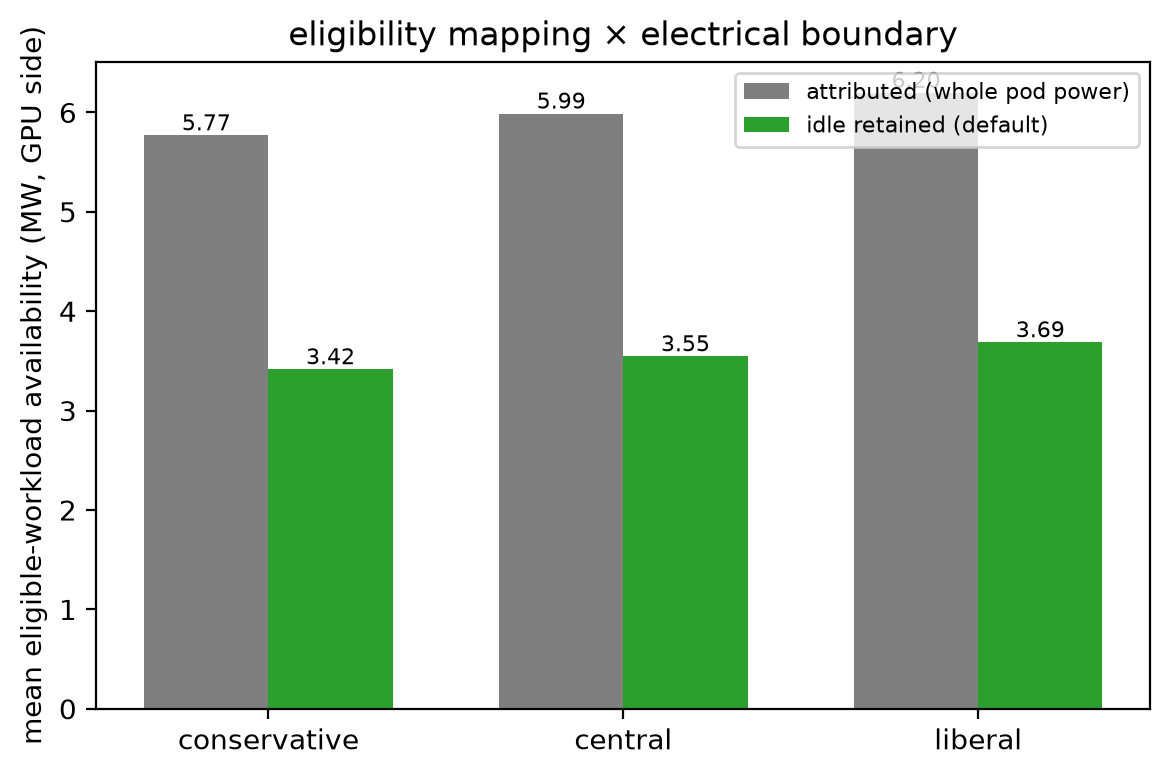}
  \end{minipage}
  \caption{Workload-semantic layers and the power boundary. (a) Mean clock-hour envelope under the central mapping. (b) Attributed low-priority workload is an upper bound; the main analysis counts only active power removed while the reclaimed GPU keeps drawing its idle floor. Both bars are workload-side, before PUE. The latter averages 3.55~\MW{} and 12.08\% of workload power.}
  \Description{The left panel stacks floor, shift, low-priority eligible, and standby workload power over a typical day. The right panel compares the larger attributed low-priority layer with the smaller active-power quantity removable when GPU idle power remains.}
  \label{fig:envelope}
\end{figure*}

The four semantic layers describe workload composition, not immediate grid relief. Under the central mapping, the floor, shift, eligible-attributed, and standby layers hold 46.12\%, 33.11\%, 20.39\%, and 0.39\% of 29.37~\MW{} mean workload power. The 20.39\% layer would remove 5.99~\MW{} only if preemption also removed the allocated GPU's idle draw. Retaining that draw lowers immediate eligible power to 3.55~\MW{}, with a 2.71~\MW{} hourly P10 (Fig.~\ref{fig:envelope}). This is 12.08\% of workload power, 6.78\% of the 52.32~\MW{} central facility series, and 6.35\% of the 55.83~\MW{} Monte Carlo median facility load.

Eligibility uncertainty changes the share by percentage points rather than multiples. The 300-draw power-model interval for idle-retained eligibility is 10.79--13.46\% of workload power. Conservative and liberal semantic mappings give 11.63\% and 12.56\%, and measured-H100 re-anchoring gives 10.38--13.22\%. Eligible power across the 15 clusters that carry it has a median pairwise correlation of 0.227. The floor varies from 43.6\% to 49.8\% across 30-day windows rather than following a monotone trend, and recorded Standby first appears on day~109, which may reflect a schema change rather than the start of standby operation. A month-one percentage cannot encode the full six-month envelope.

\subsection{The observed scheduler adds no dependable delay product}\label{sec:scheduler}

Scheduling displacement is measurable only after the execution join becomes complete enough. Over days 110--184, matched allocated-GPU-pod power averages 31.24~\MW{} on the GPU side before PUE. The actual-minus-arrival difference has positive and negative 99th-percentile magnitudes of 0.245 and 0.226~\MW{}, or 0.783\% and 0.723\% of that matched GPU power. The median daily energy swing is 0.827~\MWh{}, 0.110\% of daily matched GPU energy. These figures describe shifting the production scheduler already exercised, not an operator-controlled battery.

Newly arriving work with observed delay supplies essentially no callable capacity. Replaying only high-priority training and offline-inference arrivals that waited at least one hour yields 0.00794~\MW{} on average, 0.083\% of the 9.61~\MW{} shift layer in the post-coverage window. Its hourly P10 and $K_{0.95,1}$ are both zero; the four-hour result is smaller. The often tempting operation ``shift-layer power times fraction delayed'' gives 0.19~\MW{}, but it assigns a class-average delay rate to running work and is not a feasible backlog. We therefore exclude delay-based shifting from every capacity headline.

\subsection{Duration and reliability defeat a single percentage}\label{sec:surface}

\begin{figure*}[t]
  \centering
  \begin{minipage}[t]{0.52\textwidth}
    \centering\includegraphics[width=\linewidth]{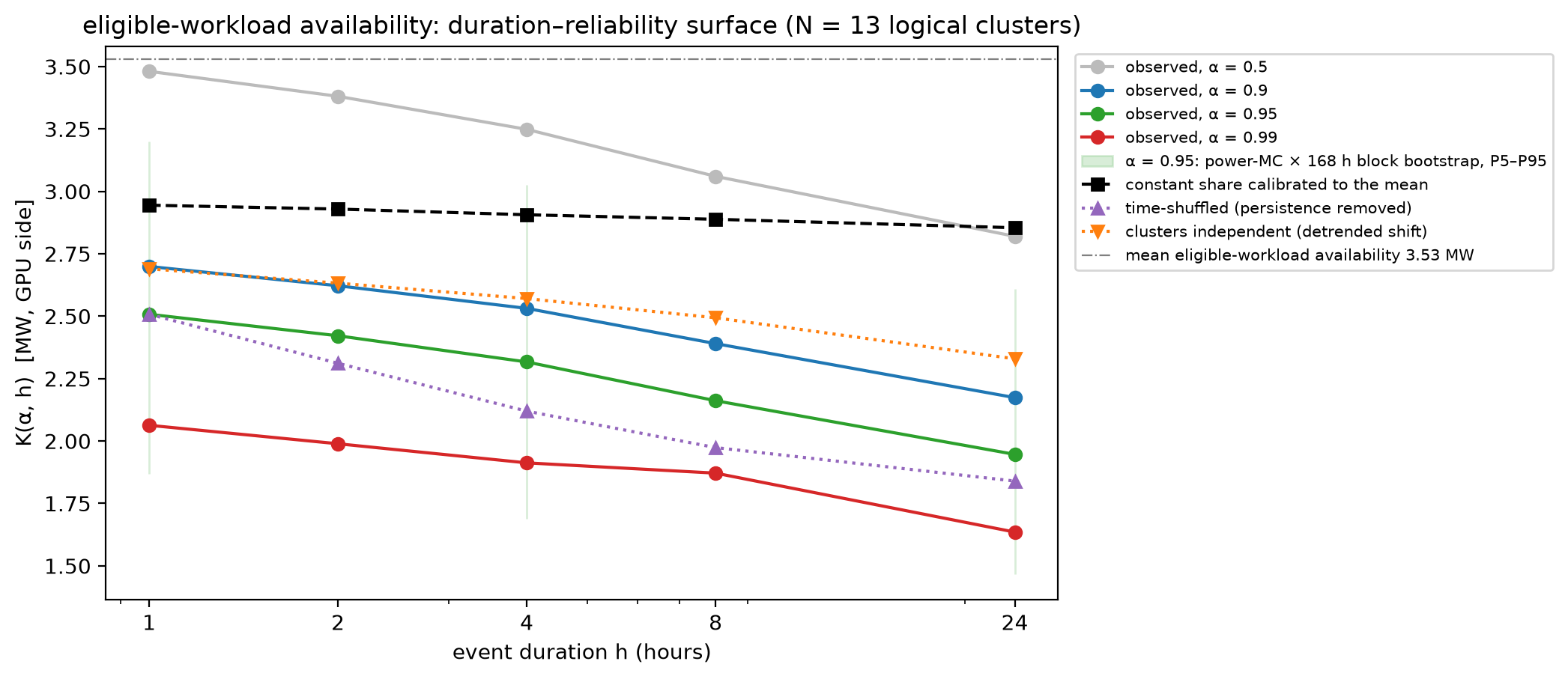}
  \end{minipage}\hfill
  \begin{minipage}[t]{0.46\textwidth}
    \centering\includegraphics[width=\linewidth]{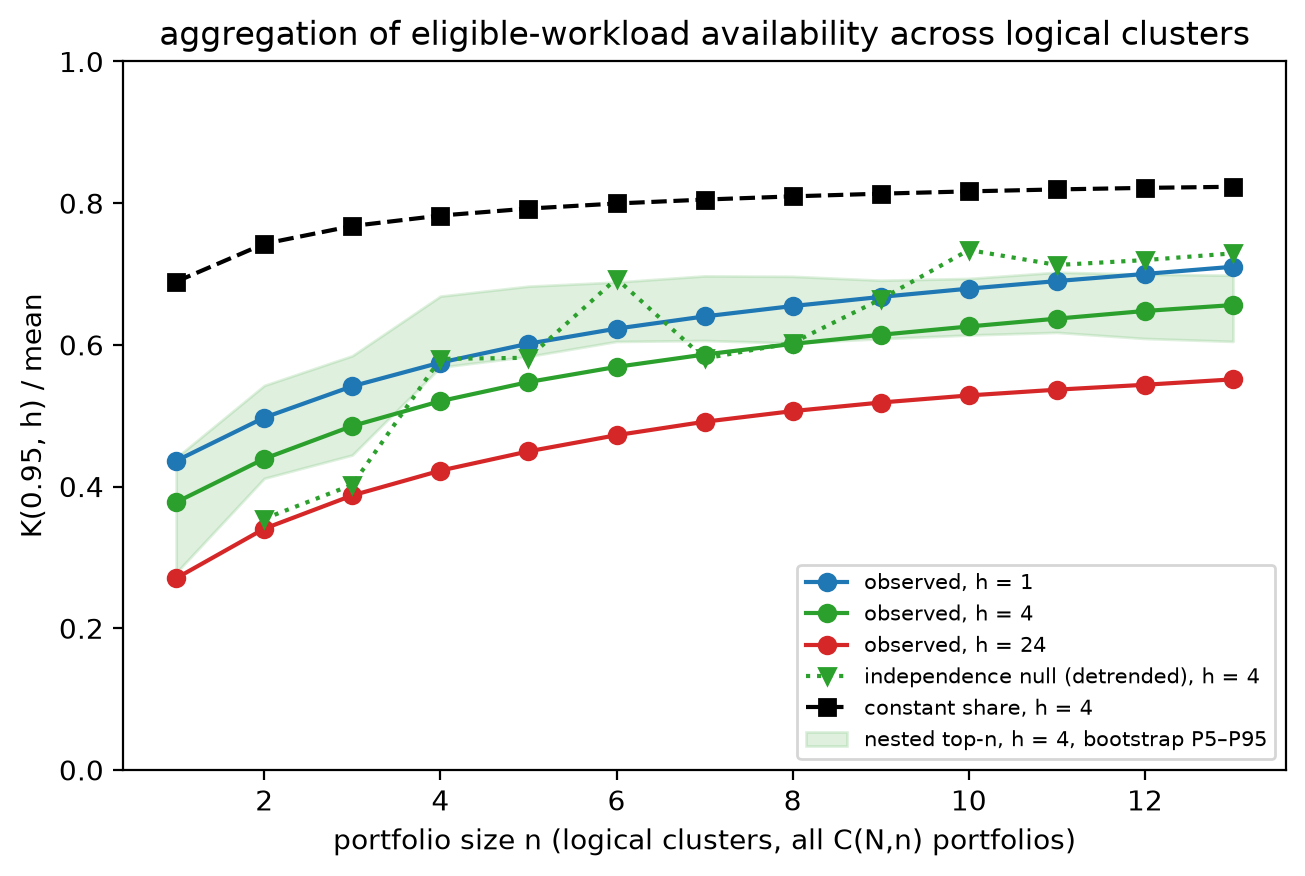}
  \end{minipage}
  \caption{Dependable eligible-workload availability at full realization ($q=1$). (a) $K_{\alpha,h}$ against duration, with the mean-share, time-shuffled, and independent-cluster counterfactuals. (b) Four-hour, 95\% firmness against portfolio size. These are historical eligibility quantities, not verified controlled response.}
  \Description{The left panel shows dependable megawatts declining with event duration and reliability, with scalar and independence counterfactuals above the observed curve. The right panel shows firmness increasing with the number of clusters but remaining below an independent-cluster benchmark.}
  \label{fig:surface}
\end{figure*}

Historical eligible-workload availability falls as the requested product becomes stricter. Across the 13-cluster portfolio, mean eligible power is 3.53~\MW{}. Under $q=1$, $K_{0.95,h}$ is 2.51, 2.42, 2.32, 2.16, and 1.95~\MW{} for $h=1,2,4,8,$ and 24 hours; firmness falls from 0.710 to 0.551 (Fig.~\ref{fig:surface}a and Table~\ref{tab:surface}). The paired 168-hour block-bootstrap and power-model interval for the four-hour value is 1.69--3.03~\MW{}. A 99\% four-hour estimate has only 44 overlapping empirical tail windows and belongs in Appendix~\ref{app:surface} rather than a headline.

A scalar fails in both directions because its calibration point fixes the error. Multiplying load by mean eligibility overstates $K_{0.95,h}$ by 17.4\%, 25.5\%, and 46.6\% at one, four, and 24 hours; the paired interval for the four-hour ratio is 22.0--37.4\%. Tail calibration to the observed four-hour product instead makes the one-hour product 6.4\% low, matches four hours by construction, and makes the 24-hour product 16.9\% high. No one scalar matches the surface across durations. More generally, the workload share $p_{h,\alpha}=K_C(h,\alpha)/K_L(h,\alpha)$ that reproduces each point of the surface, where $K_L$ is the same running-minimum quantile of workload power, ranges from 0.98 of the mean eligible share at one hour and 50\% availability to 0.61 at 24 hours and 99\% (Table~\ref{tab:required}); at 95\% it is 0.85, 0.79, and 0.68 of the mean share for one, four, and 24 hours, and no single share is within 10\% of all three. Assuming clusters are independent adds 7.3\%, 11.0\%, and 19.7\% at the same horizons, while time-shuffling understates the four-hour value by 8.5\% because it destroys persistence, although that shuffle also removes seasonality and growth.

\begin{table}[t]
\caption{Workload share $p_{h,\alpha}=K_C(h,\alpha)/K_L(h,\alpha)$ that reproduces each point of the availability surface, as a fraction of the mean eligible share (12.08\%). Fleet series of the 15 eligible clusters at $q=1$.}
\label{tab:required}
\centering
\small
\begin{tabular}{lcccc}
\toprule
$h$ & $\alpha=0.5$ & 0.9 & 0.95 & 0.99 \\
\midrule
1 hour & 0.98 & 0.88 & 0.85 & 0.74 \\
2 hours & 0.96 & 0.85 & 0.83 & 0.72 \\
4 hours & 0.93 & 0.83 & 0.79 & 0.70 \\
8 hours & 0.89 & 0.79 & 0.74 & 0.70 \\
24 hours & 0.83 & 0.73 & 0.68 & 0.61 \\
\bottomrule
\end{tabular}
\end{table}

Portfolio diversification is valuable but saturates below independence. Mean four-hour firmness rises from 0.378 for one cluster to 0.656 for all 13, a 1.74-fold gain (Fig.~\ref{fig:surface}b). The independent-cluster counterfactual reaches 0.729. This difference connects the covariance result in Section~\ref{sec:aggregation} to a planning quantity: ignoring co-movement grants roughly one additional tenth of $K_{0.95,4}$.

\begin{table*}[t]
\caption{Fleet eligible-workload availability and scalar errors at $\alpha=0.95$ and $q=1$. Ratios divide each counterfactual $K$ by observed $K$; values above one overstate availability. The mean is 3.53~\MW{} across the 13 included clusters. Figure~\ref{fig:surface}a and Appendix~\ref{app:surface} report the other reliability levels.}
\label{tab:surface}
\centering
\begin{tabular}{lrrrrr}
\toprule
Duration & Observed $K$ (MW) & Firmness & Mean-share scalar & $K_4$ tail-calibrated & Independent clusters \\
\midrule
1 hour  & 2.508 & 0.710 & 1.174 & 0.936 & 1.073 \\
4 hours & 2.317 & 0.656 & 1.255 & 1.000 & 1.110 \\
24 hours& 1.947 & 0.551 & 1.466 & 1.169 & 1.197 \\
\bottomrule
\end{tabular}
\end{table*}

\subsection{Realizability and recalibration bound the megawatt claim}\label{sec:realizability}

An unmeasured response fraction changes the level but not the shape of the result. Because Eq.~\ref{eq:k} is homogeneous, four-hour, 95\%-available relief for the 13-cluster portfolio is $2.3168q$~\MW{}; at least $q=0.432$ is needed to offer 1~\MW{}, and $q=0.5$ yields 1.16~\MW{}. The duration and portfolio comparisons survive any common $q$, but the absolute megawatt claim does not. Field preemption, checkpoint, service-level, and telemetry tests are needed to estimate $q$.

Rolling-origin validation separates historical fit from an operational offer. With eight training weeks and four test weeks, mean coverage before derating is 0.954, 0.946, and 0.907 for one-, four-, and 24-hour products targeted at 0.95. A consistent absolute-MW calibration needs no downward adjustment at one hour, a 1.3\% adjustment at four hours, and an 11.5\% adjustment at 24 hours, yielding 2.29 and 1.73~\MW{} for the latter two products. Longer commitments are less stable under fleet evolution even when the full-history surface looks smooth. Start time matters less than drift: across the 24 trace-local start hours the four-hour 95\% value ranges from 2.13 to 2.53~\MW{} (median 2.34, lowest for starts at hour~21), across the seven phases of the weekly cycle from 2.14 to 2.55~\MW{}, and across 30-day windows from 1.90 to 2.77~\MW{}. Grid calls are not uniform in time, so a product covering specific hours should use the conditional values.

Load relief is a power service rather than an energy saving. If every deferred megawatt-hour rebounds and the expected lost work equals half of a one-hour checkpoint interval, net energy change after a four-hour call is $-12.5\%$ of gross curtailed energy. For the full-realization 2.32~\MW{} offer, that rebound contains 10.44~\MWh{}. The mean noneligible workload power, defined as the floor plus shift layers, is 23.264~\MW{}; absorbing the rebound at 10\% of that level takes 4.49 hours. This rule implies 8.49 hours between four-hour calls and at most 19.8 calls per week. Appendix~\ref{app:surface} reports the full realizability, recovery, and rolling-origin sensitivities.
 \section{Related Work}\label{sec:related}

Power modeling work spans device telemetry and site-level synthesis, while our study begins from scheduler records. The Fan--Weber--Barroso model established utilization as a server-power predictor~\cite{fan2007power}, and later fleet models reported low aggregate error from richer counters~\cite{radovanovic2021powermodel}. GPU studies now measure training, inference, idle, and execution-idle states at device or node scale~\cite{patel2024polca,patel2024splitwise,newkirk2025h100,vercellino2026nlr,vadari2026parking,chung2025mlenergy,enskat2026mfu,execidle2026}. Compositional generators learn sub-minute inference-state transitions and can preserve energy and autocorrelation under new serving configurations~\cite{wilkins2026servers}. Our hourly reconstruction gives up device-scale dynamics to recover six months of workload type, priority, cluster hierarchy, and scheduling history. The two scales answer different grid questions: device traces test ramp and control speed, while the production trace measures prevalence, persistence, and portfolio covariance.

Scheduler-aware studies establish algorithms and demonstrations but rarely expose a production-wide denominator. Carbon-aware and virtual-capacity systems move compute across time or sites~\cite{radovanovic2022carbon,hall2024vcc}. A controlled 256-GPU demonstration verified multi-hour response~\cite{colangelo2025nature}, and Acun et al. measured GPU power-capping capability before scaling the result through simulation~\cite{acun2026flex}. Trace-based work closest to ours estimates delay flexibility from an older single-cluster Alibaba trace~\cite{caprara2026ladflex} or uses HPC and request traces to calibrate a workload-composition model~\cite{majumder2026composition}. Data-center demand response has a long computing-side literature~\cite{wierman2014opportunities}, including fleet-wide power capping in production~\cite{wu2016dynamo}, and deep-learning cluster schedulers routinely preempt, checkpoint, and resize jobs~\cite{xiao2018gandiva,gu2019tiresias,qiao2021pollux}, which is the mechanism behind the priority labels we use. Our contribution is neither a new scheduler nor a response controller. It measures how much production work enters each semantic class, then preserves its serial and cross-cluster structure when translating eligibility to dependable megawatts.

Grid-planning studies show why this empirical input matters. Flexible load can reduce capacity, cost, or congestion, but the result depends strongly on location, event shape, and the assumed division among fixed, shiftable, and interruptible tiers~\cite{khanal2026shift,chen2026defer,kim2026siting}. Duke's screening study illustrates the other endpoint by treating a large load as fully curtailable for a limited number of hours~\cite{norris2025rethinking}. Those studies solve broader power-system problems than ours; we test the input abstraction they must choose. Treating 20\% and 100\% of central facility load as curtailable produces facility-side capacities 3.5--4.5 and 18--22 times the trace-derived $K_{0.95,h}$, respectively, at $q=1$ under the main one-to-one marginal workload-to-facility conversion (2.9--5.2 and 15--26 times over the four-hour bootstrap band; Table~\ref{tab:formulations} in Appendix~\ref{app:surface}). These tiers are modelling assumptions, not measurements; the comparison states what they imply on this trace, not that the cited studies err. A generic percentage can therefore dominate the downstream system result even when the optimization model is exact.

\section{Discussion and Limitations}\label{sec:discussion}

The duration--reliability surface is a better contract primitive than a flexibility share. An interconnection study can select an event duration and target availability, apply a verified realizable fraction $q$, and derate the estimate under rolling calibration. A portfolio operator can then claim only the diversification observed across its clusters rather than the larger independent-site credit. This representation is still compact: five durations, four reliability levels, and one response factor replace a single percentage while exposing the assumptions that change a capacity result.

Eligibility is valuable evidence but not verified response. Low-priority status shows that an operator already accepts preemption semantics, yet the trace does not reveal checkpoint completion, service-level violations, restart energy, network bottlenecks, or the control latency required by a grid product. The idle-retained boundary avoids the largest accounting error by leaving GPU idle power in place, and the $q$ analysis makes the remaining implementation gap explicit. A field trial should estimate $q$ by workload class and horizon, measure host and cooling response, and record rebound. Until then, $K_{\alpha,h}(S;1)$ is dependable historical eligibility and an upper bound on delivered power.

Portfolio covariance deserves the same attention as individual-site eligibility. The trace shows complementarity among workload types inside a cluster but positive residual covariance among clusters. Calendar removal and capacity normalization account for only 34.4\% of the excess, and stable-capacity plateaus rule out expansion as the dominant mechanism. Common submissions, model releases, maintenance, or unobserved scheduler state are plausible explanations, but the public data cannot distinguish them. This unidentified covariance still changes the planning answer: the independent-cluster counterfactual overstates four-hour relief by 11\% and 24-hour relief by 20\%.

The MISO South calculation in Appendix~\ref{app:interconnection} is an illustrative peak-cap screen rather than a network study. With a 5\% margin and a 0.5\%-of-hours response budget, the trace-derived envelope raises screened headroom from 1.684 to 1.809~GW under the conservative marginal-response convention, while 20\% and fully curtailable representations give 2.105 and 3.054~GW. The mean-share scalar gives 1.807~GW, only 0.15\% below the trace result. This near equality is specific to the alignment-and-cap screen: the case tests the residual-load floor and the danger of generic 20\% or 100\% assumptions, not the duration claim. Transmission, reserves, contingencies, forecast error, and generator deliverability remain outside the screen.

The single-operator trace bounds external validity. Workload labels, priority policy, hardware mix, and cluster coordination may differ at another operator or even in the next six months of the same fleet. The monthly envelope and rolling-origin results show this drift directly. We therefore release per-cluster series and mappings rather than presenting 12.1\% as a universal AI-data-center constant. Replication across operators matters more than fitting a richer model to the same 185 days.

Hourly utilization and modeled power bound temporal validity. The trace cannot test sub-hour response time, short training oscillations, or minimum power during a 15-minute settlement interval. Public 5-second node traces suggest that a 15-minute minimum is often 91--97\% of the hourly mean, but one LoRA case has a five-minute P10 of 0.48; these measurements are suggestive rather than a fleet correction. Applying a concave curve to hourly mean utilization can also overstate training power by 22--25\% under an extreme two-state duty cycle, whereas the linear family is unbiased. The host model, 31\% of facility power, is a linear per-core model whose parameters enter the sensitivity draws but are not independently validated. Power-meter calibration at the traced facility would narrow all of these uncertainties.

Semantic mapping and missing joins bound construct validity. Central eligibility treats every low-priority pod as preemptible because the source system gives that class a short graceful-preemption path, but the trace does not record whether a given pod actually checkpoints successfully. Conservative and liberal mappings move the eligible share between 11.63\% and 12.56\%, and power re-anchoring widens it to 10.38--13.22\%. Only 47.1\% of spans and 58.8\% of GPU-hours match delay records over the full trace, and the abrupt coverage change forces the post-day-110 restriction. Within that restricted window, however, a direct audit matches 99.99996\% of GPU-hours, so unjoined work cannot explain the 0.008~\MW{} backlog. The result is strong evidence that observed delay did not create a dependable product in that window, not proof that the workloads could never be shifted under a different policy.
 
\section{Artifact Availability}\label{sec:artifact}
The raw Alibaba cluster-trace-gpu-v2026 is distributed by its authors through the Alibaba ClusterData repository and linked object storage~\cite{asi_trace_2026}; we do not redistribute it. Regional demand in Appendix~\ref{app:interconnection} comes from the US Energy Information Administration's Hourly Electric Grid Monitor~\cite{eia930}. The derived data products are public in the tagged release \texttt{v1.0.1} of \url{https://github.com/meiyilutaustin/asi-trace-power}, archived at Zenodo under \url{https://doi.org/10.5281/zenodo.22308423}: the 858,816-row sufficient-statistics table (attached to the GitHub release; the Zenodo archive holds the source snapshot), the hourly fleet and cluster power bands, the flexibility envelopes on all three electrical boundaries, the $K_{\alpha,h}$ tables for every portfolio and era, the compiled power-measurement table, and the source data and result tables behind every figure and quoted number (\texttt{results/headline.json}). Derived products are released under CC BY-NC 4.0 subject to the research-use condition of the source trace.
 The complete analysis pipeline is public at \url{https://github.com/meiyilutaustin/asi-trace-power} (MIT license; tag \texttt{v1.0.1}, archived as \url{https://doi.org/10.5281/zenodo.22308423}, is the exact version used for the reported runs): raw-trace aggregation into sufficient statistics, the workload-conditioned power model with the online-inference floor and the three preemption boundaries, the Monte Carlo driver, the envelope and dependable-availability analyses with their paired counterfactuals, the execution-span extraction and scheduler counterfactuals, the external-validation and Jensen-bound scripts, the interconnection screen, the realizable-fraction, rolling-origin and covariance-decomposition robustness checks, and the Slurm job files used for the reported runs. A stage-to-figure map in the repository README identifies the script behind each result.
 
\section{Conclusion}\label{sec:conclusion}

A hyperscale production trace turns data-center flexibility from a percentage into a measurable time series. The 155,410-GPU fleet contains 12.1\% immediate eligible workload power under an idle-retained preemption boundary, yet only 6.35\% of median facility demand. Its 95\%-available quantity falls from 2.51~\MW{} at one hour to 1.95~\MW{} at 24 hours under full realization, and a mean-calibrated scalar overstates the latter by 47\%. Portfolio aggregation improves firmness, but positive cross-cluster covariance makes the independence assumption optimistic. Observed scheduling delay contributes zero 95\%-available backlog capacity.

The practical result is a contract specification rather than a universal flexibility factor. Duration, reliability, cluster portfolio, realizable response fraction, recovery, and recalibration each change the megawatts that can be offered. Publishing those dimensions alongside the underlying envelope lets grid planners use production evidence without treating eligibility as proven control. The next experimental step is small and concrete: preempt representative low-priority workloads, meter the facility boundary, and estimate $q$ and rebound by horizon.
 
\bibliographystyle{ACM-Reference-Format}
\bibliography{main}

\appendix
\section{Power Model and Anchor Comparison}\label{app:power}

\subsection{Central parameters and uncertainty ranges}

The power model uses public accelerator ratings and measured workload anchors. Table~\ref{tab:anchors} separates parameters fixed by the trace from parameters varied in Monte Carlo. Unknown XPU buckets draw one rating per bucket and replicate, so model uncertainty remains correlated across the fleet. The online-inference floor is expressed as total GPU power divided by TDP and converted internally to the active span above idle.

\begin{table*}[t]
\caption{Central power parameters and Monte Carlo ranges. Curve entries are fractions of the idle-to-TDP span at utilization knots $\{0,0.25,0.5,0.75,1\}$.}
\label{tab:anchors}
\centering
\begin{tabular}{lll}
\toprule
Parameter & Central value & Monte Carlo treatment \\
\midrule
A10 / A30 / L20 rating & 150 / 165 / 275 W & independent model multiplier $U(0.85,1.15)$ \\
A100 / A800 rating & 400 W & $U(300,460)$ W \\
H800 rating & 700 W & model multiplier $U(0.85,1.15)$ \\
H20 / H20-141GB rating & 400 W & model multiplier $U(0.85,1.15)$ \\
XPU-A through XPU-E & 475 W & $U(250,700)$ W \\
Allocated-GPU idle fraction $\phi$ & 0.20 & $U(0.15,0.25)$ \\
Online-inference floor & 0.43 TDP & $U(0.35,0.50)$ TDP \\
Training curve & (0,.50,.75,.90,1.00) & anchored or linear family \\
Online-inference curve & (0,.20,.35,.45,.55) & anchored or linear family \\
Offline-inference curve & (0,.30,.50,.70,.85) & anchored or linear family \\
Null-utilization curve value & 0.30 & $U(0.10,0.50)$ \\
Host idle / peak & 1.5 / 4.0 W per core & $U(1,2)$ / $U(3,6)$ \\
PUE & 1.2 & $U(1.1,1.4)$ \\
\bottomrule
\end{tabular}
\end{table*}

\begin{figure*}[t]
  \centering
  \includegraphics[width=0.82\textwidth]{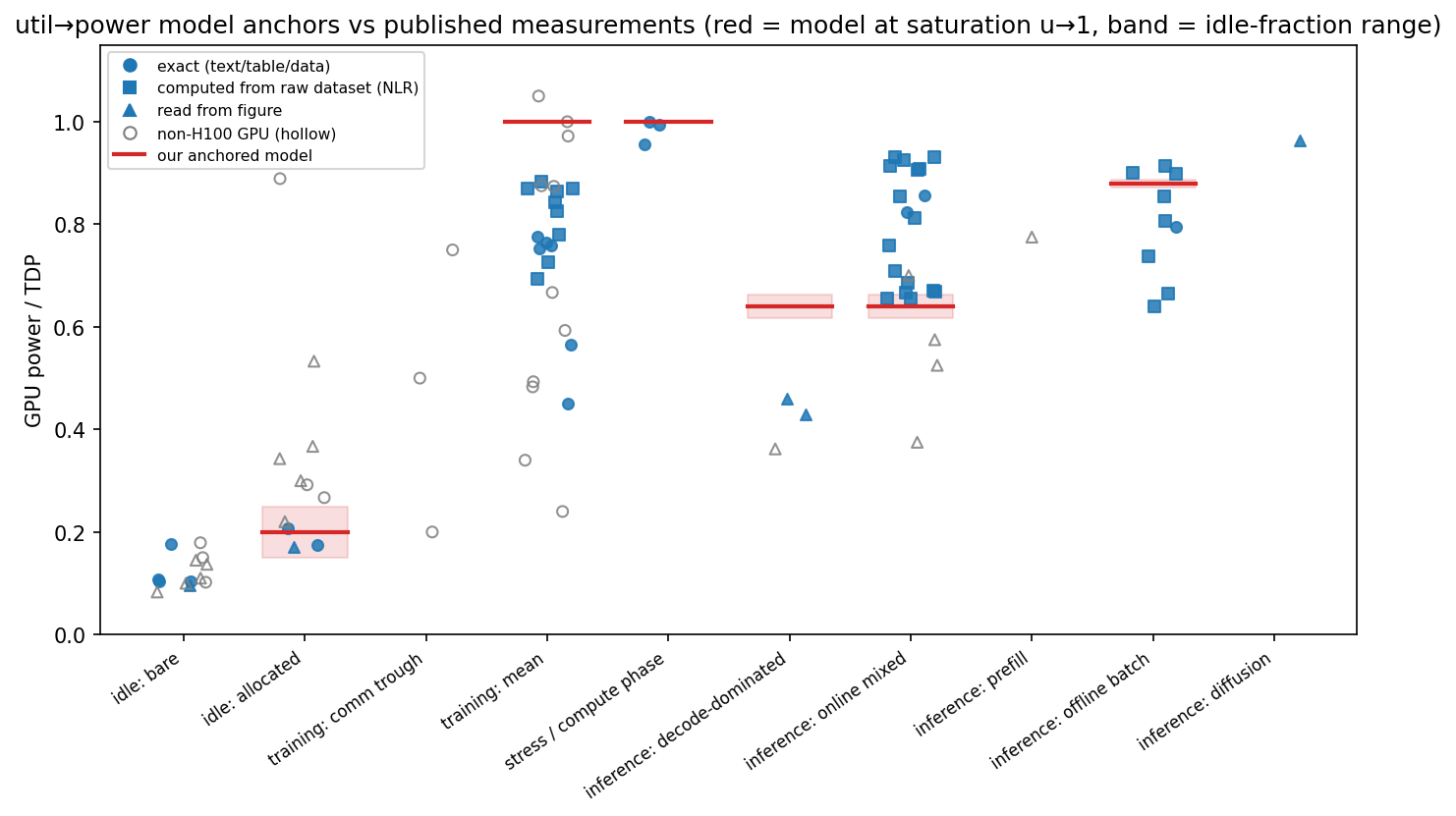}
  \caption{Power-model anchors against 92 categorized public measurements. Markers report power as a fraction of device TDP; horizontal model marks show central category values. The compilation includes raw-log calculations, reported tables, and approximate figure readings with provenance tags.}
  \Description{A categorical scatter plot compares many measured fractions of GPU rated power for idle, training, and inference workloads with horizontal model anchors.}
  \label{fig:anchors}
\end{figure*}

The published measurements support the scale of the anchors but not a facility-specific calibration. The collection contains 117 points tagged as exact, recomputed, or figure-read; 92 belong to categories represented in the model (Fig.~\ref{fig:anchors}). Allocated idle observations include the 0.20 setting, online-inference observations span 0.36--0.93 of TDP around the 0.43 floor and 0.64 saturation value, and offline batch inference spans 0.64--0.91 around the 0.88 setting. Mean training observations span roughly 0.69--0.87 in the closest H100 studies, so a curve that reaches 1.0 can be 15--45\% high relative to those means~\cite{vercellino2026nlr,newkirk2025h100,patel2024polca,patel2024splitwise,chung2025mlenergy,enskat2026mfu,vadari2026parking,execidle2026}.

\begin{figure*}[t]
  \centering
  \begin{minipage}[t]{0.49\textwidth}
    \centering\includegraphics[width=\linewidth]{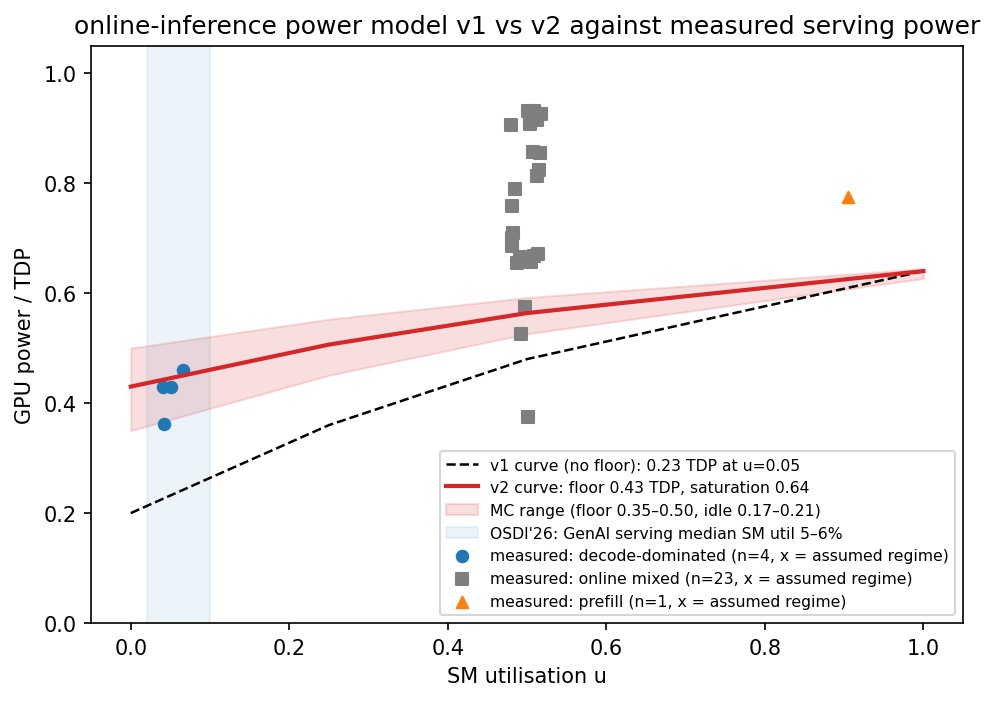}
    \caption*{(a) Online-inference floor.}
  \end{minipage}\hfill
  \begin{minipage}[t]{0.49\textwidth}
    \centering\includegraphics[width=\linewidth]{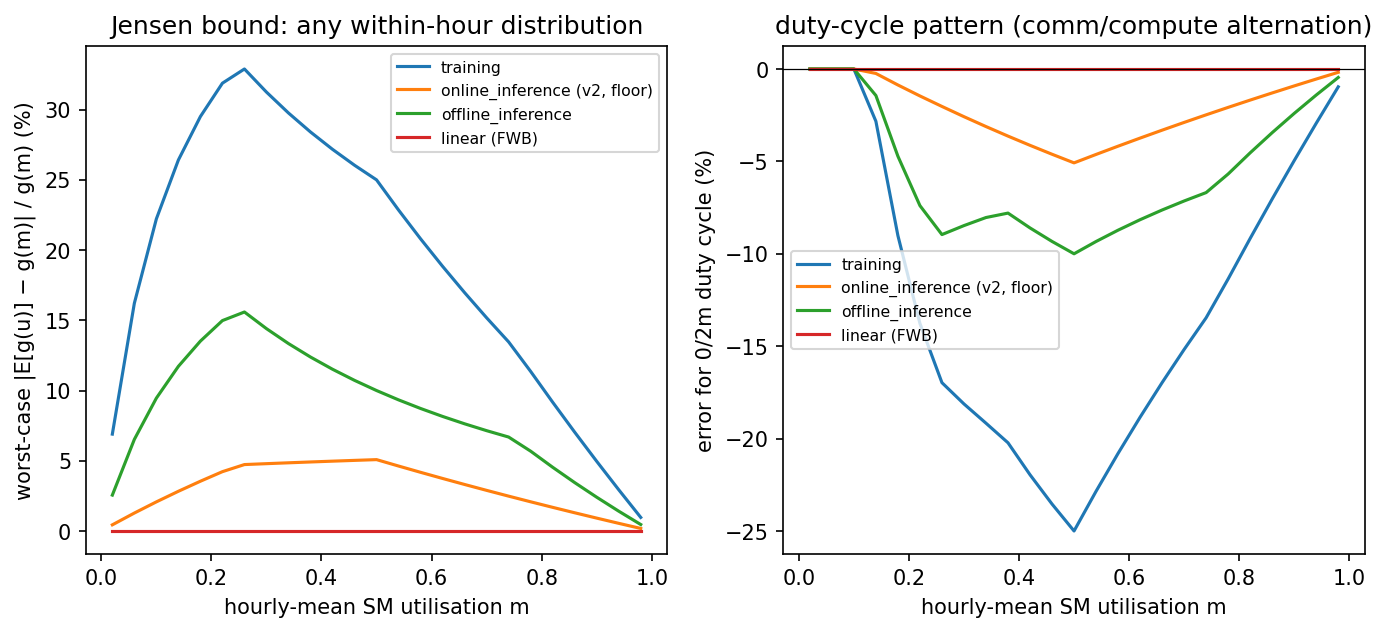}
    \caption*{(b) Hourly concavity bound.}
  \end{minipage}\\[3pt]
  \includegraphics[width=0.70\textwidth]{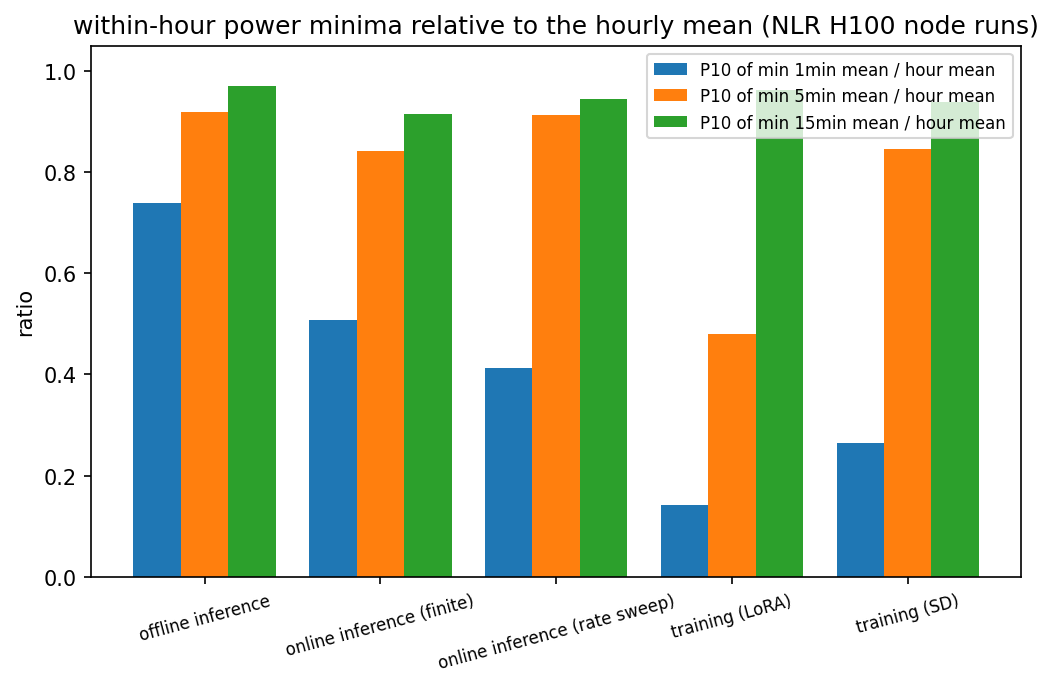}
  \caption*{(c) Sub-hour power minima.}
  \caption{Power-model boundary and temporal checks. (a) Online-inference curve before and after adding the 0.43-TDP total-power floor, with measured serving observations. (b) Jensen error from evaluating a piecewise-linear curve at hourly mean utilization; negative duty-cycle error means the reconstruction is high. (c) Minimum one-, five-, and 15-minute power divided by the hourly mean in public 5-second node traces.}
  \Description{The first panel compares online-inference power curves with measured observations. The second shows the hourly-mean reconstruction error for several two-state workload patterns. The third compares minimum sub-hour power fractions for inference and training workloads.}
  \label{fig:validationdetails}
\end{figure*}

A nonzero online floor corrects a structural failure in curves forced through the origin. Decode-dominated inference can draw about 0.43 of TDP even when reported SM utilization is low, so the final curve raises the intercept and compresses its remaining active span to preserve saturation (Fig.~\ref{fig:validationdetails}a). This change raises the must-run floor and removes low-utilization online inference from the apparent curtailment pool.

Hourly averaging creates a one-sided risk for concave curves. If utilization alternates between zero and twice its hourly mean, applying the training curve to the mean overstates average power by 22--25\% over representative means; the linear family remains unbiased (Fig.~\ref{fig:validationdetails}b). This calculation is a stress bound, not an empirical error distribution, and the Monte Carlo family mixture does not prove coverage of within-hour behavior.

Re-anchoring changes the level but preserves the capacity conclusion. The measured-mid case produces 54.49~\MW{} median facility power and 13.22\% idle-retained eligibility; the measured-low case produces 52.82~\MW{} and 10.38\%; the base produces 55.83~\MW{} and 12.08\% (Table~\ref{tab:reanchor}). The workload-power denominator changes with the curve anchors, so the table reports it separately rather than holding the base denominator fixed. All three place immediate eligibility near one tenth of workload power rather than the 20.39\% attributed layer.

\begin{table}[t]
\caption{Sensitivity to measured-H100 curve anchors. Facility power is the Monte Carlo median; eligibility is idle-retained power divided by workload power.}
\label{tab:reanchor}
\centering
\begin{tabular}{lrrr}
\toprule
 & Base & Mid & Low \\
\midrule
Facility power (MW) & 55.83 & 54.49 & 52.82 \\
Workload power (MW) & 29.37 & 27.89 & 25.18 \\
Eligible power (MW) & 3.55 & 3.69 & 2.61 \\
Eligible share (\%) & 12.08 & 13.22 & 10.38 \\
\bottomrule
\end{tabular}
\end{table}

Public node traces suggest that settlement granularity cannot be inferred from hourly means. Across offline inference, mixed online inference, rate-sweep online inference, LoRA training, and diffusion training, the 15-minute minimum is typically 91--97\% of its hourly mean (Fig.~\ref{fig:validationdetails}c). Shorter windows vary more: LoRA has a five-minute P10 of 0.48 and a one-minute P10 of 0.14. These ratios come from our re-analysis of single-node 5-second logs~\cite{vercellino2026nlr}; they motivate a fleet-metering experiment but do not justify a universal derating.

\FloatBarrier
\section{Load and Aggregation Details}\label{app:load}

\begin{figure*}[t]
  \centering
  \begin{minipage}[t]{0.49\textwidth}
    \centering\includegraphics[width=\linewidth]{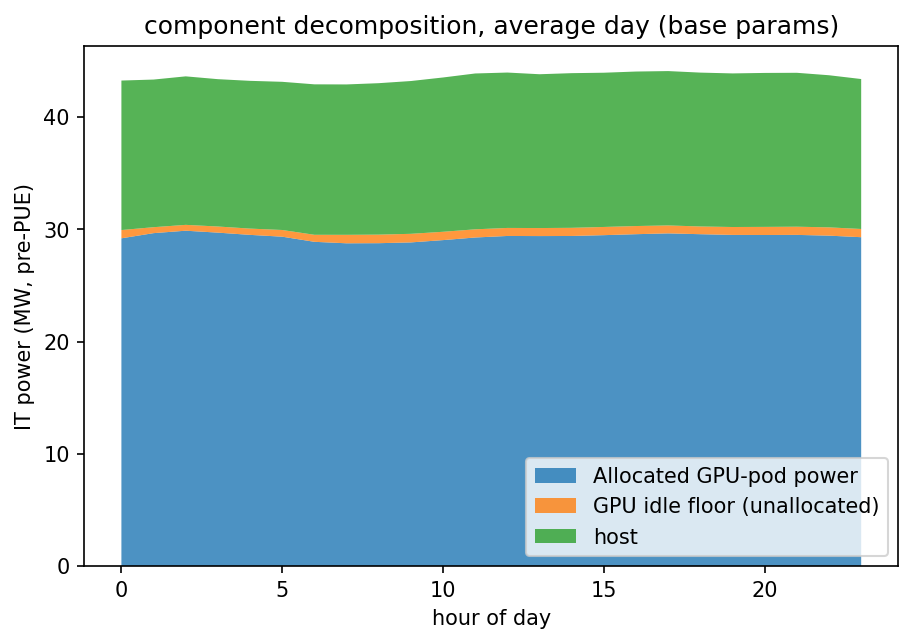}
    \caption*{(a) IT-power components before PUE.}
  \end{minipage}\hfill
  \begin{minipage}[t]{0.49\textwidth}
    \centering\includegraphics[width=\linewidth]{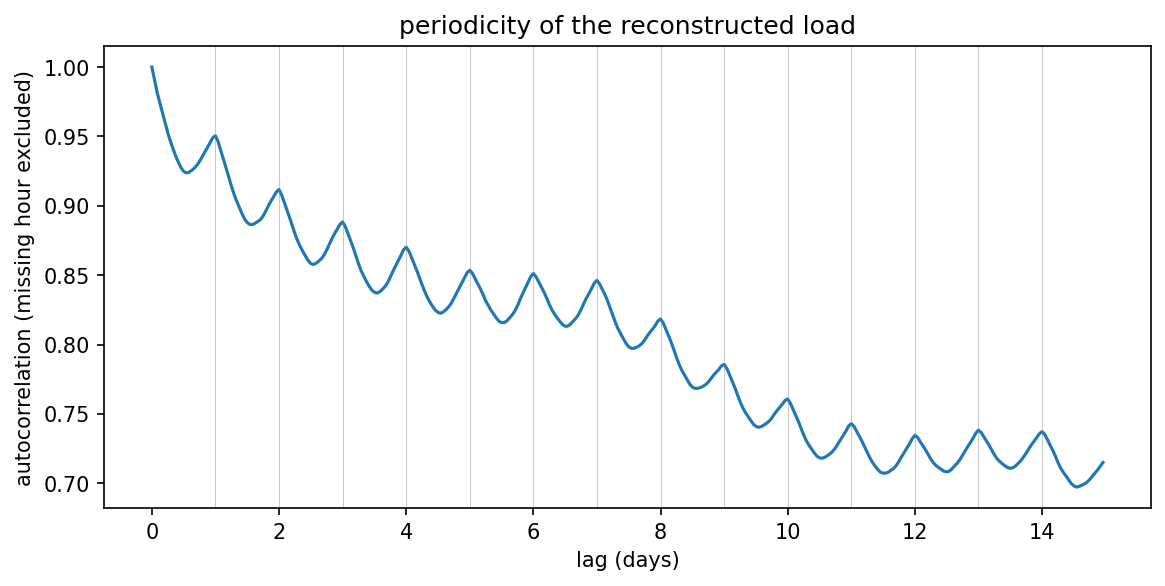}
    \caption*{(b) Raw autocorrelation.}
  \end{minipage}
  \caption{Additional load-profile diagnostics. Autocorrelation is computed on the full growing series and is reported as persistence, not as a detrended periodicity estimate.}
  \Description{A stacked area plot decomposes IT power into allocated GPU-pod power, unallocated-GPU idle power, and host power. A second plot shows autocorrelation with markers at daily and weekly lags.}
  \label{fig:components}
\end{figure*}

The load decomposition confirms that near-full allocation creates a floor. Allocated GPU-pod power contributes 67.35\% of central IT power, host power 31.07\%, and unallocated-GPU idle 1.58\% (Fig.~\ref{fig:components}a); applying one PUE preserves these shares in the central facility series. Raw autocorrelation stays high through 24-hour and seven-day lags (Fig.~\ref{fig:components}b), but the growth steps prevent a causal seasonal interpretation.

\begin{figure*}[t]
  \centering
  \includegraphics[width=0.58\textwidth]{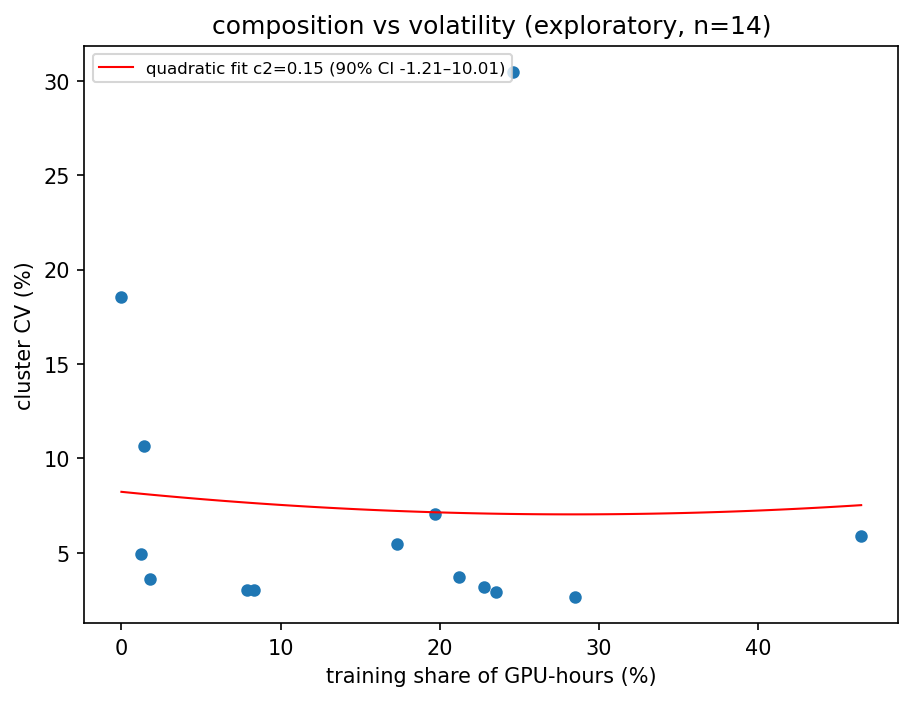}
  \caption{Exploratory cluster variability against training share. The quadratic curvature interval crosses zero; the curve is descriptive only.}
  \Description{Fourteen cluster points show coefficient of variation versus training share with a shallow fitted curve and a wide confidence band.}
  \label{fig:ushape}
\end{figure*}

The exploratory composition curve is not statistically resolved. A quadratic fit of cluster coefficient of variation against training share has curvature 0.149, but its 90\% bootstrap interval is $[-1.207,10.007]$ (Fig.~\ref{fig:ushape}). The trace therefore does not confirm the simulated U-shaped relation in prior work~\cite{majumder2026composition}.

\begin{figure*}[t]
  \centering
  \includegraphics[width=0.88\textwidth]{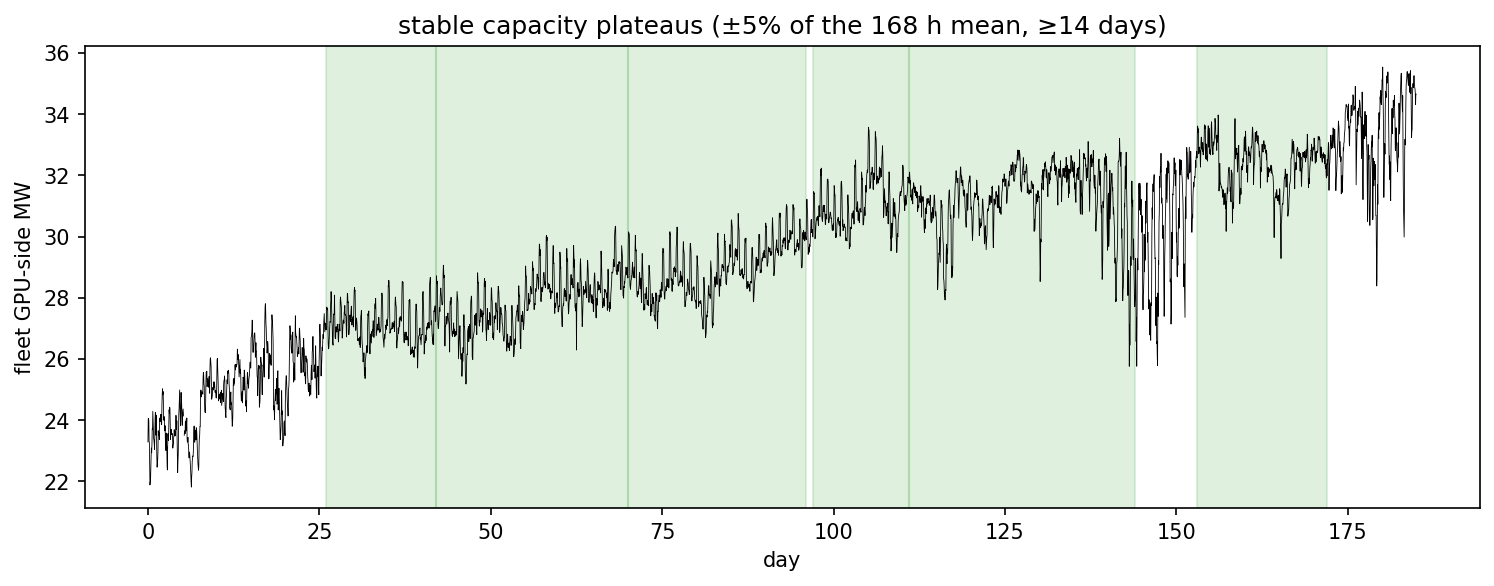}
  \caption{Cross-cluster synchrony within six stable-capacity plateaus. Raw ratios remain close to the full-window value after excluding capacity ramps.}
  \Description{A time series identifies six stable-capacity intervals and annotates their raw, calendar-adjusted, and capacity-normalized synchrony ratios.}
  \label{fig:plateaus}
\end{figure*}

The stable-capacity test rules out expansion as the dominant covariance mechanism. Across six plateaus, raw cluster-to-fleet synchrony ranges from 1.645 to 1.900 and averages 1.765 (Fig.~\ref{fig:plateaus}). Capacity normalization within those same plateaus leaves a 1.567 average ratio. The result supports a portfolio correction while leaving its operational cause open.

\section{Scheduler and Envelope Details}\label{app:scheduler}

\begin{figure*}[t]
  \centering
  \begin{minipage}[t]{0.59\textwidth}
    \centering\includegraphics[width=\linewidth]{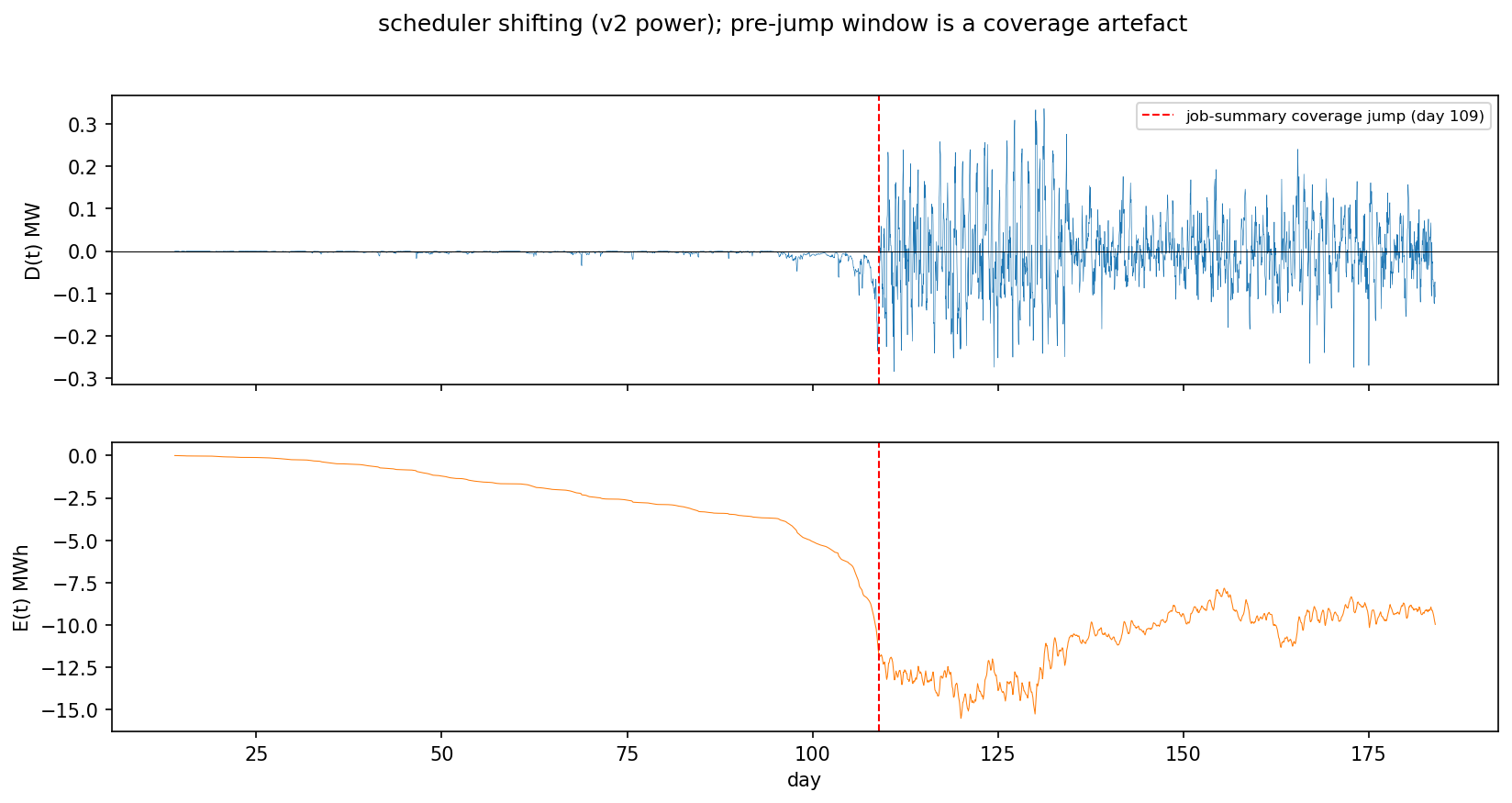}
  \end{minipage}\hfill
  \begin{minipage}[t]{0.39\textwidth}
    \centering\includegraphics[width=\linewidth]{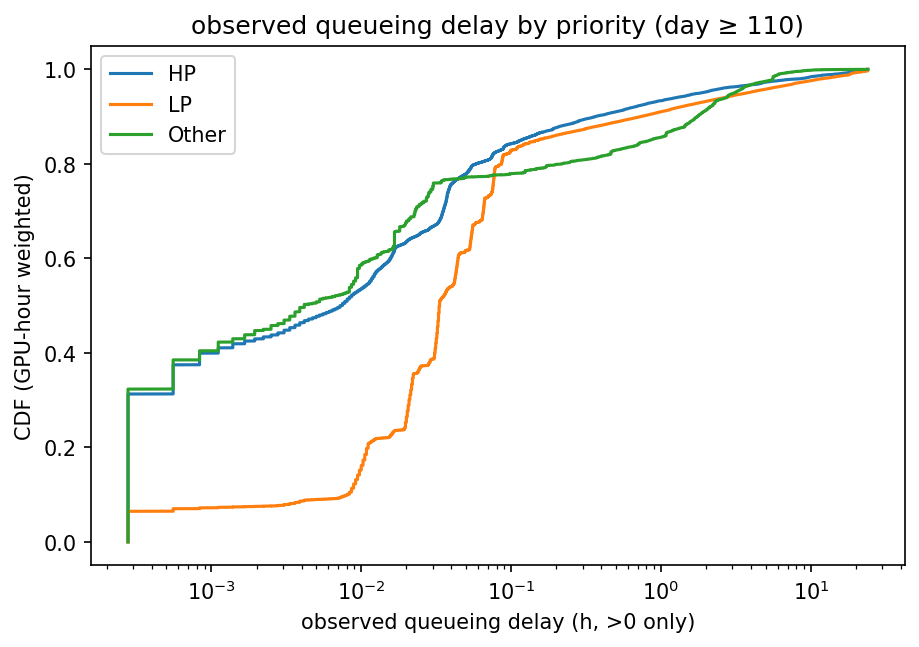}
  \end{minipage}
  \caption{Observed scheduler displacement after the execution-join coverage stabilizes. (a) Difference power and integrated energy. (b) GPU-hour-weighted scheduling-delay distributions by priority. Delay is an observed outcome, not an exogenous tolerance label.}
  \Description{The left panel plots small positive and negative scheduling-displacement power and its cumulative energy after day 110. The right panel shows delay survival curves split by priority.}
  \label{fig:scheduler}
\end{figure*}

The delay-derived difference process becomes interpretable only after the record-coverage jump. Figure~\ref{fig:scheduler} shows actual minus arrival-time counterfactual allocated-GPU-pod power and its integral over days 110--184, alongside the observed delay distribution. A direct join audit in this window finds 251,882,189.67 matched and 108.58 unmatched GPU-hours, for 99.9999569\% coverage; the unmatched mass spans only three pod-days. Although 113,940 of 36,685,018 execution-summary pod identifiers carry conflicting workload or priority labels, even assigning every unmatched GPU-hour to one class bounds composition bias at 0.0000431\% of window GPU-hours. On this GPU-side, pre-PUE boundary, the post-coverage P99 discharge and charge are 0.245 and 0.226~\MW{}, and the median daily cycle is 0.827~\MWh{}. These values measure displacement already present under the recorded scheduler.

\begin{figure*}[t]
  \centering
  \includegraphics[width=0.84\textwidth]{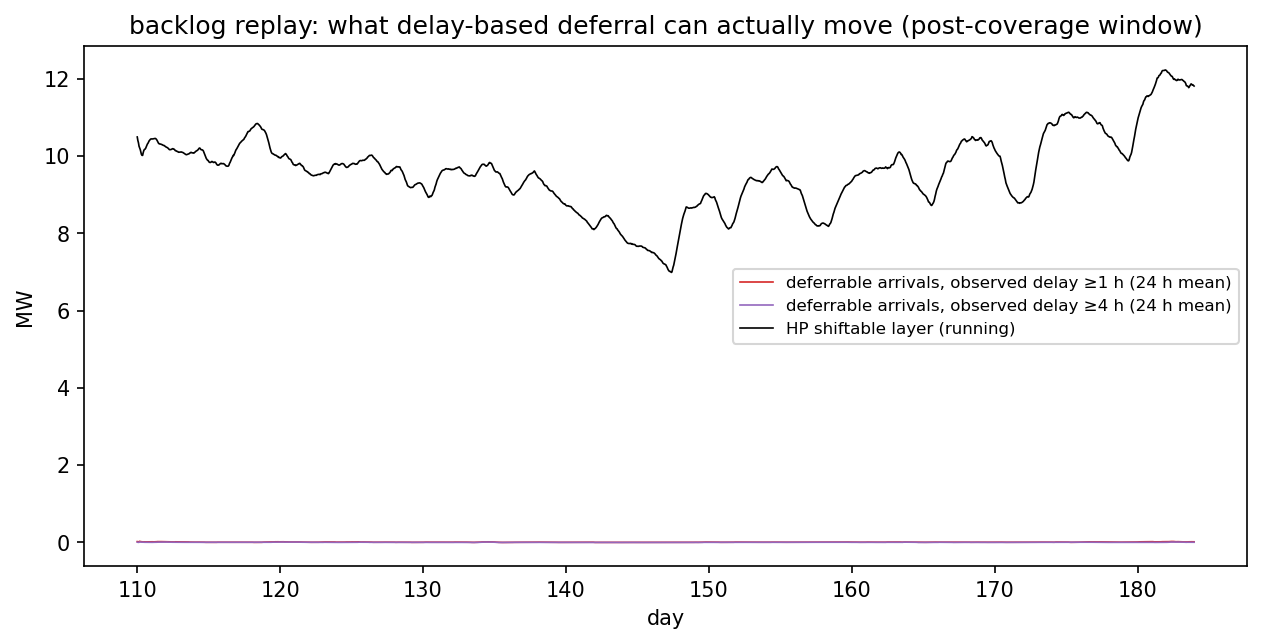}
  \caption{Newly deferrable high-priority arrivals under the observed-delay replay. The available backlog is close to zero compared with the semantic shift layer.}
  \Description{A time series and distribution compare a very small newly deferrable arrival backlog with a much larger semantic shiftable-workload layer.}
  \label{fig:backlog}
\end{figure*}

The feasible-arrival replay is much smaller than the semantic shift layer. One-hour eligible arrivals average 0.00794~\MW{} and four-hour arrivals 0.00353~\MW{} after day~110, with zero P10 and zero 95\%-available capacity at both horizons (Fig.~\ref{fig:backlog}). The one-hour quantity is 0.083\% of the 9.61~\MW{} shift layer. Running work could add flexibility only through an unobserved pause or checkpoint action.

\begin{figure*}[t]
  \centering
  \includegraphics[width=0.98\textwidth]{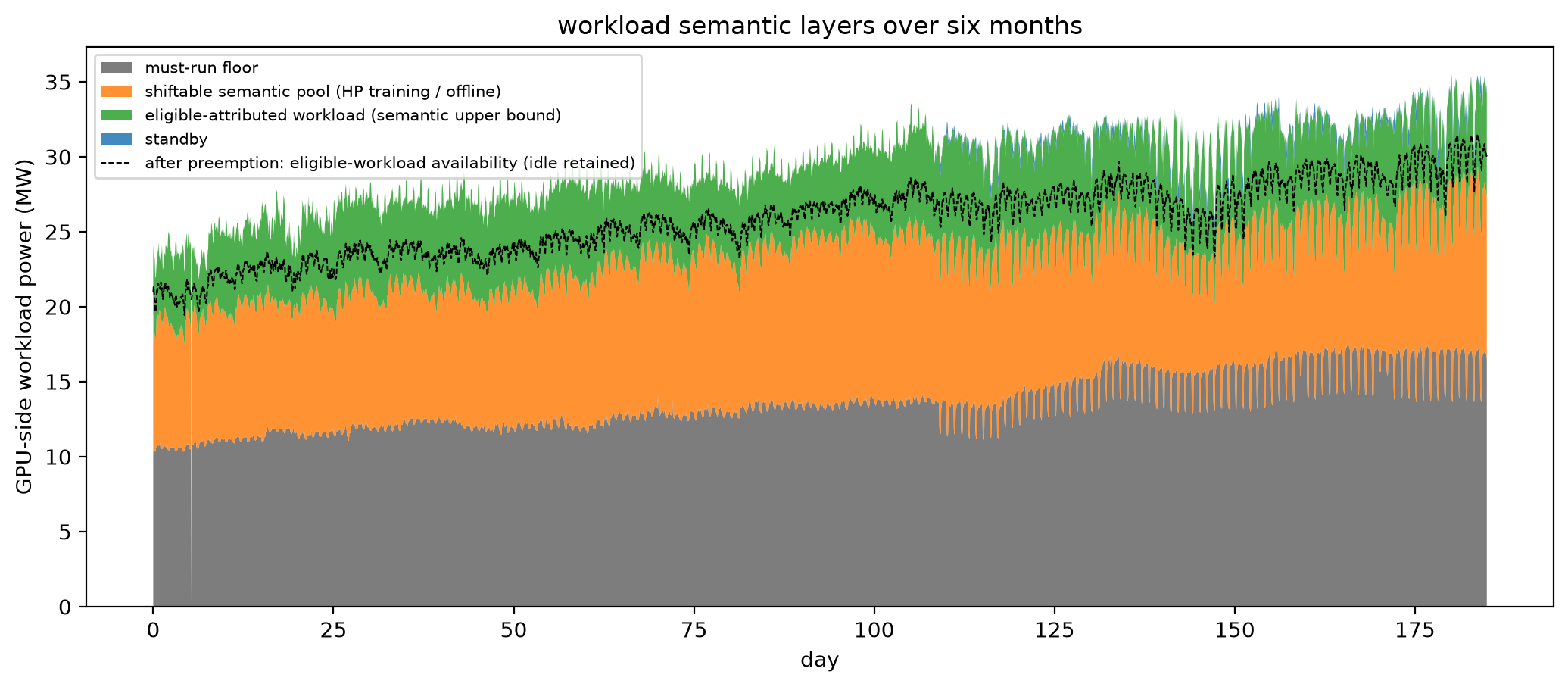}\\[2pt]
  \begin{minipage}[t]{0.49\textwidth}
    \centering\includegraphics[width=\linewidth]{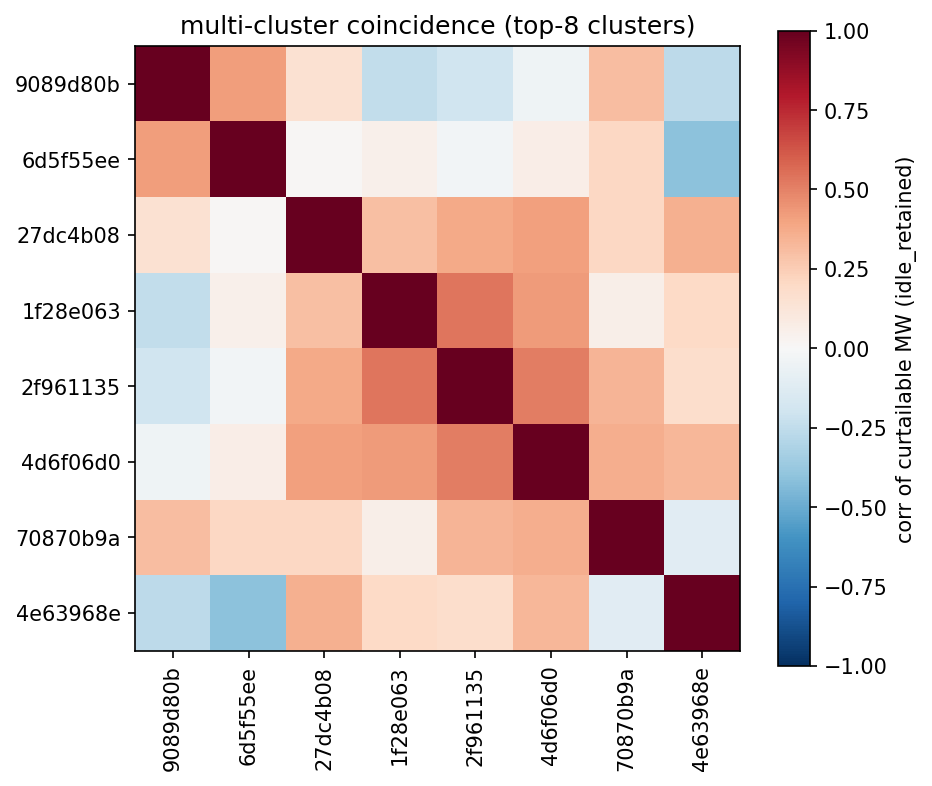}
  \end{minipage}\hfill
  \begin{minipage}[t]{0.49\textwidth}
    \centering\includegraphics[width=\linewidth]{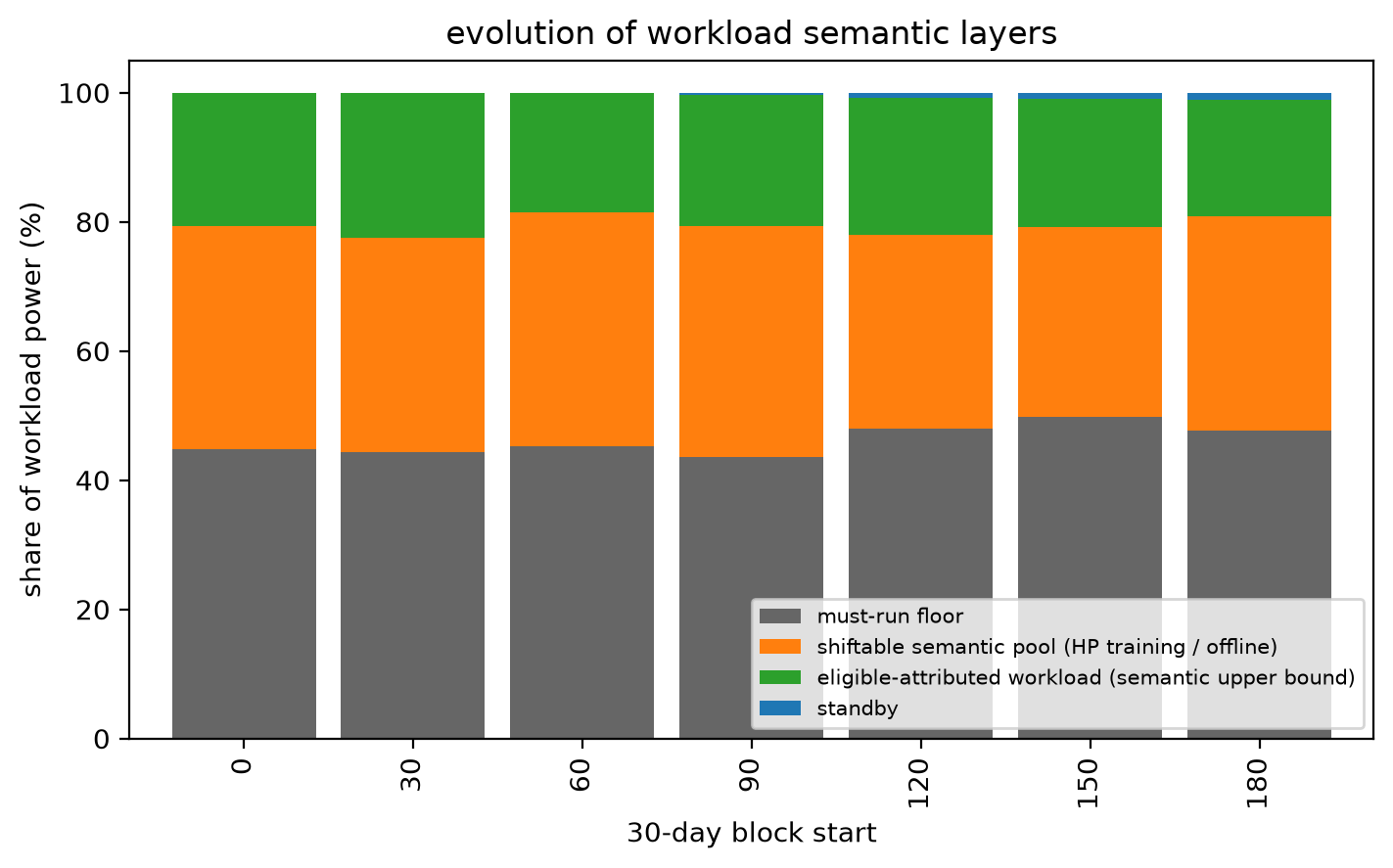}
  \end{minipage}
  \caption{Envelope evolution and portfolio structure. Top: semantic layers over 185 days; ``eligible attributed'' is the full low-priority workload layer and exceeds idle-retained relief. Bottom left: pairwise eligible-power correlations. Bottom right: 30-day layer shares.}
  \Description{A stacked time series shows floor, shift, eligible-attributed, and standby workload layers. Two smaller panels show a cluster correlation heat map and layer shares across successive monthly windows.}
  \label{fig:envelopedetails}
\end{figure*}

The hourly envelope changes in both level and composition. Figure~\ref{fig:envelopedetails} shows the complete 185-day layer series, pairwise eligible-power correlations, and 30-day layer shares. Standby first appears on day~109 and remains 1.08\% in the last window; the central floor ranges from 43.57\% to 49.78\% across windows. Eligible-power correlation has a 0.227 median across the 15 participating clusters.

\section{Dependable-Availability Robustness}\label{app:surface}

\begin{figure*}[t]
  \centering
  \begin{minipage}[t]{0.45\textwidth}
    \centering\includegraphics[width=\linewidth]{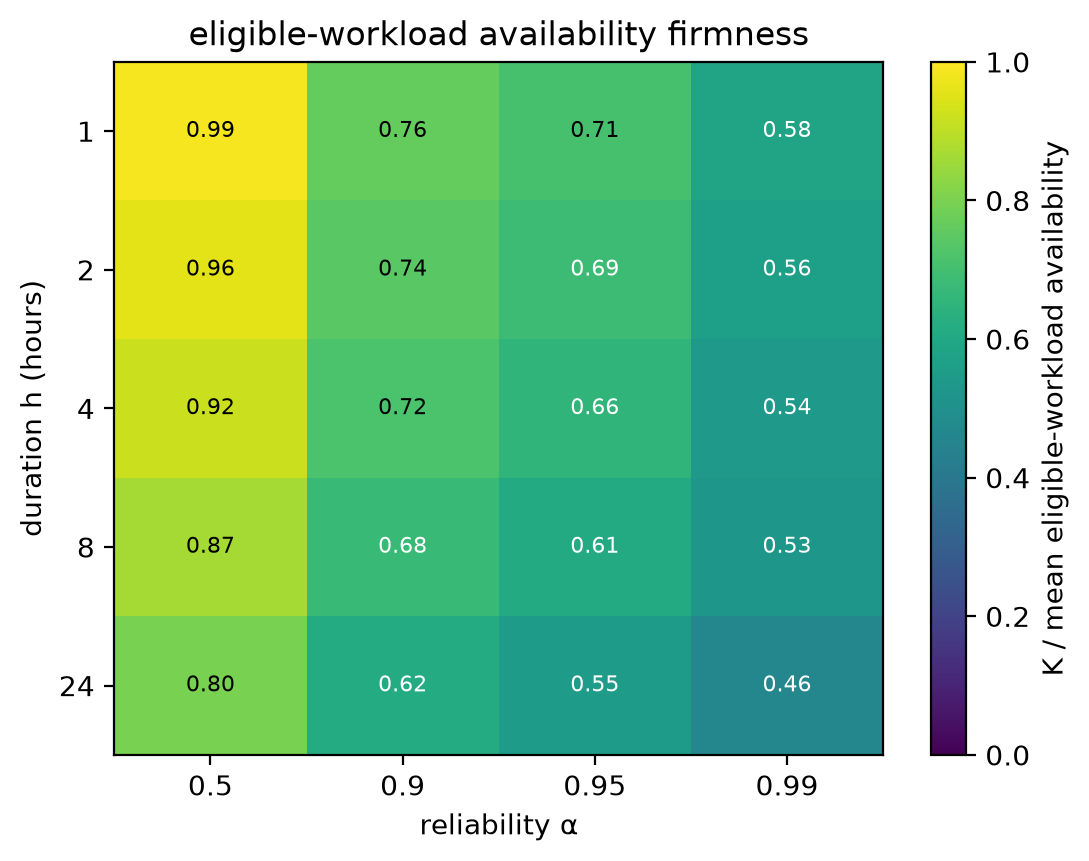}
  \end{minipage}\hfill
  \begin{minipage}[t]{0.53\textwidth}
    \centering\includegraphics[width=\linewidth]{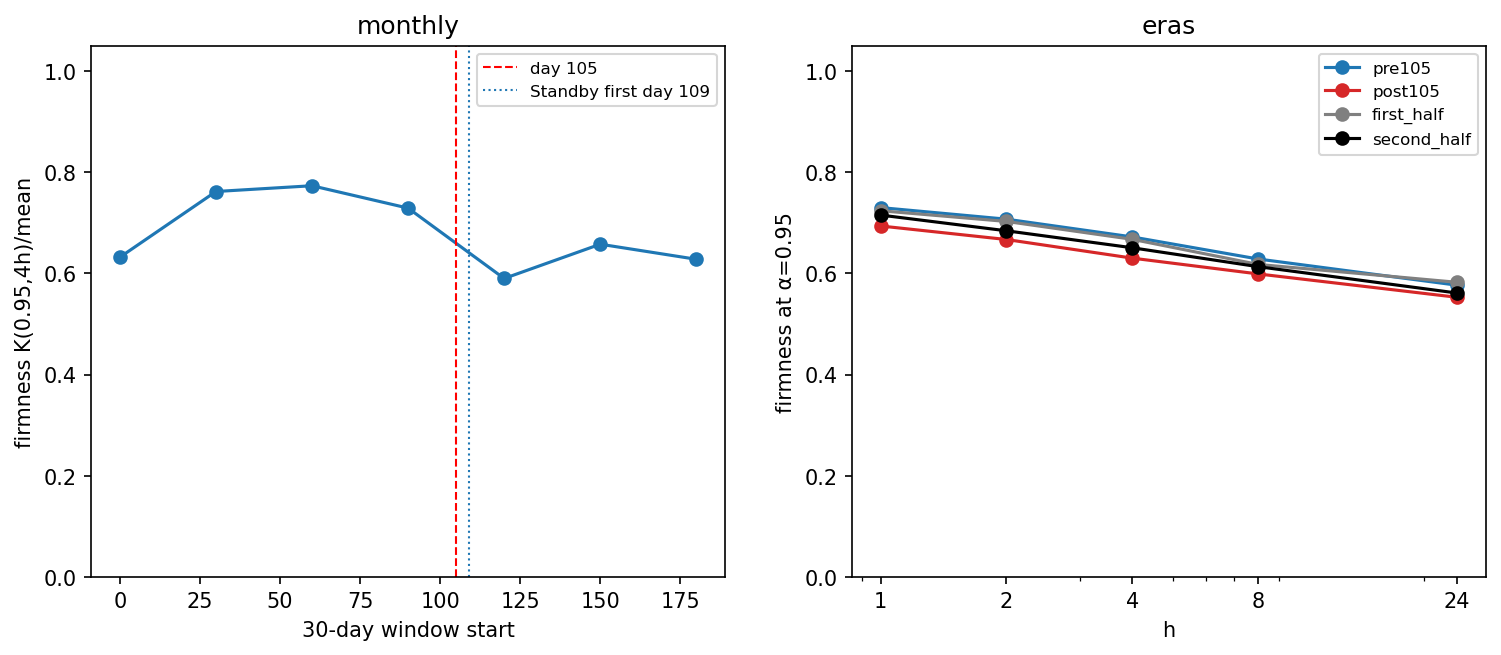}
  \end{minipage}
  \caption{Dependable-availability sensitivities at $q=1$. (a) Firmness over the duration--reliability grid. (b) Four-hour estimates across trace eras and 30-day windows.}
  \Description{A heat map shows firmness falling with longer duration and higher reliability. A second plot compares four-hour dependable availability across trace halves, periods, and monthly windows.}
  \label{fig:surfaceextra}
\end{figure*}

The full duration--reliability grid shows a smooth decline rather than a threshold. Figure~\ref{fig:surfaceextra}a reports firmness for four reliability levels and five durations; the 99\% four-hour cell is based on 44 tail windows. Figure~\ref{fig:surfaceextra}b shows estimates by trace period and month, which range from 0.59 to 0.77 in four-hour firmness as workload mix and capacity evolve.

\begin{figure*}[t]
  \centering
  \includegraphics[width=0.58\textwidth]{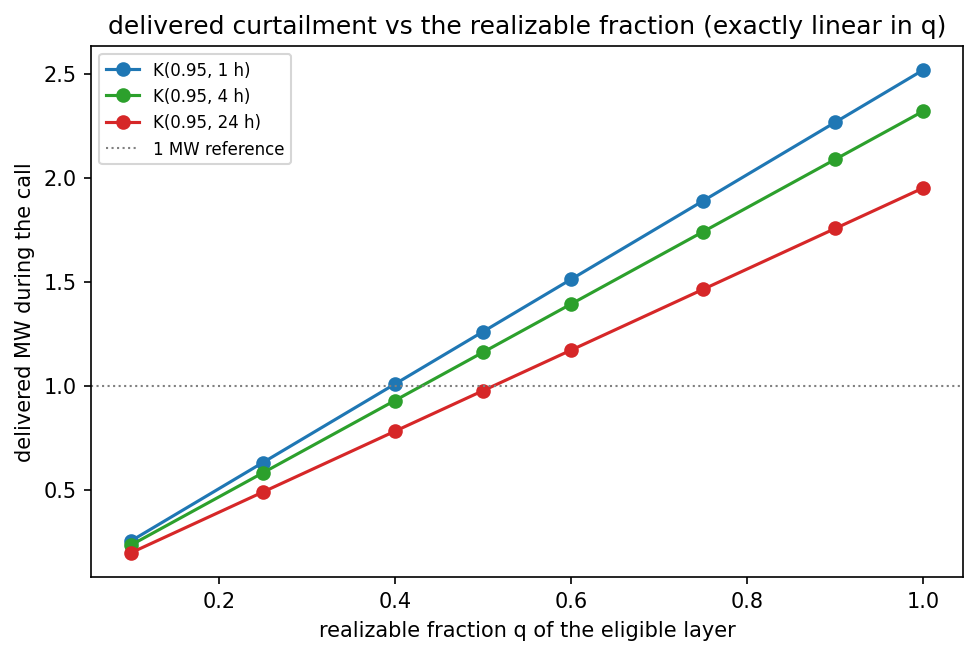}
  \caption{Full-fleet four-hour, 95\%-available power against realizable fraction $q$. The relationship is exactly linear; the main 13-cluster support-filtered portfolio differs by less than 0.2\%.}
  \Description{A straight line maps realizable fraction from zero to one to dependable megawatts from zero to about 2.32.}
  \label{fig:q}
\end{figure*}

The realizable fraction changes only the vertical scale. Figure~\ref{fig:q} verifies $K_{0.95,4}(q)=2.3199q$~\MW{} to numerical precision for the full 15-cluster eligible fleet. The support-filtered 13-cluster portfolio used for the main duration surface gives $2.3168q$~\MW{}, so its 1 and 1.8~\MW{} offers require $q\geq0.432$ and $q\geq0.777$; the full-fleet thresholds are 0.431 and 0.776. This relationship lets a field trial update the absolute offer without re-estimating duration or portfolio shape if $q$ is common across hours.

\begin{figure*}[t]
  \centering
  \begin{minipage}[t]{0.58\textwidth}
    \centering\includegraphics[width=\linewidth]{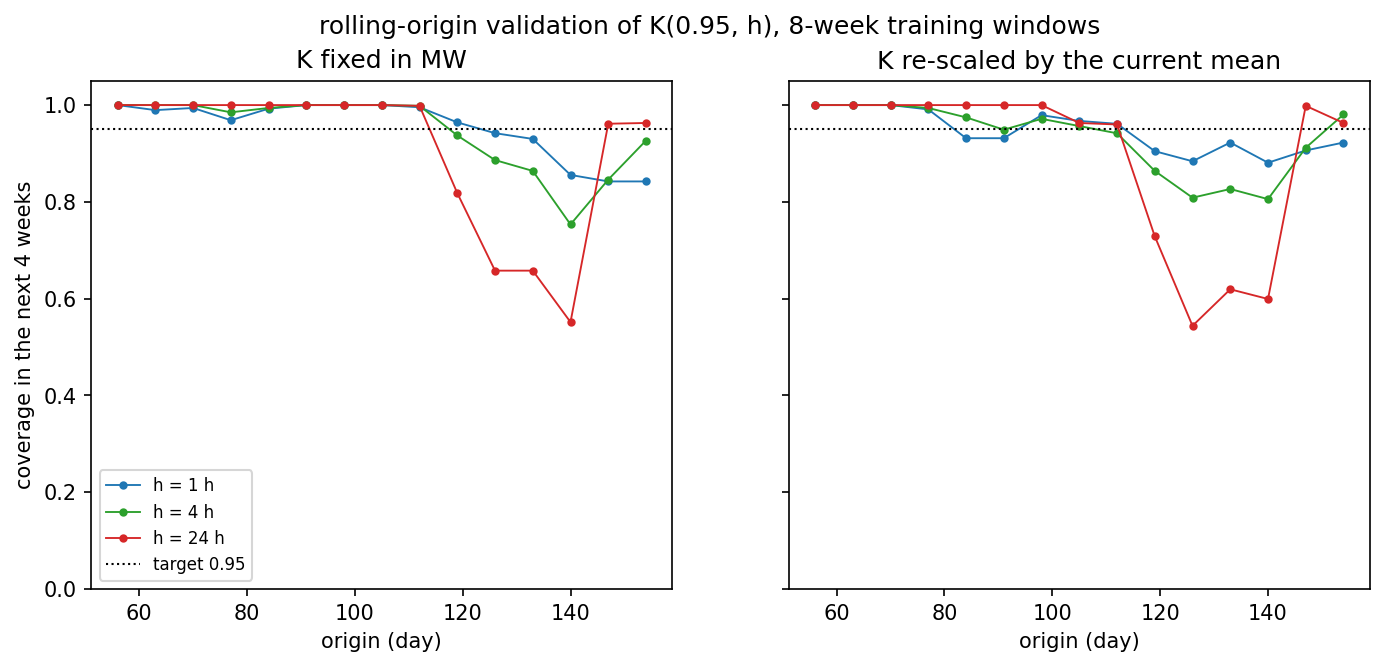}
  \end{minipage}\hfill
  \begin{minipage}[t]{0.40\textwidth}
    \centering\includegraphics[width=\linewidth]{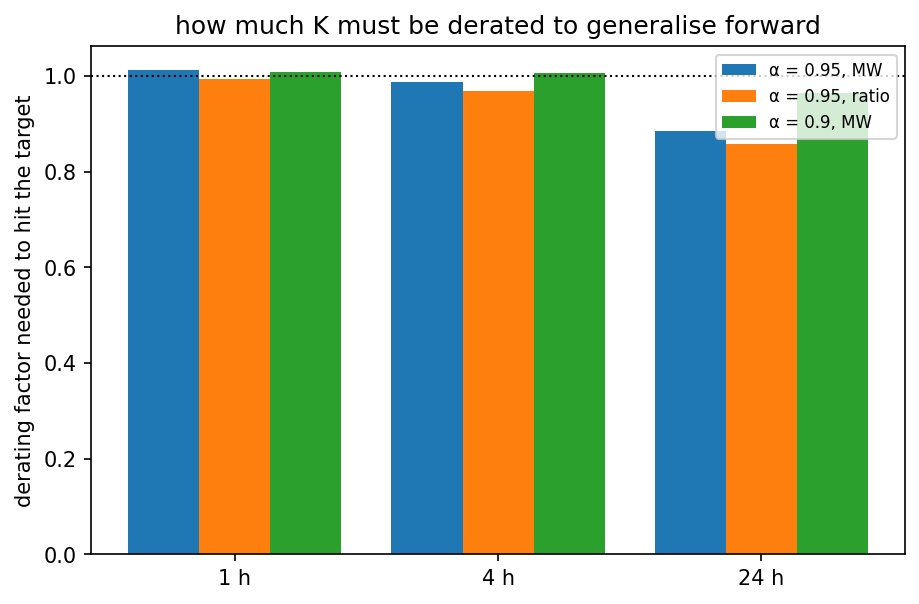}
  \end{minipage}
  \caption{Prospective rolling-origin check. (a) Test-window coverage by origin. (b) Multiplicative factors needed for target mean coverage under absolute and normalized specifications. Main-text values consistently use the absolute-MW column.}
  \Description{The left panel shows coverage varying across fifteen rolling tests, with greater shortfalls for 24-hour products. The right panel shows stronger downward adjustment for longer durations.}
  \label{fig:rolling}
\end{figure*}

Rolling-origin tests reveal horizon-dependent drift. Figure~\ref{fig:rolling} uses 15 eight-week-train, four-week-test origins. Absolute-MW mean coverage is 0.954, 0.946, and 0.907 at one, four, and 24 hours before calibration. The corresponding calibration factors are 1.011, 0.987, and 0.885, giving 2.546, 2.290, and 1.727~\MW{}; normalized-by-mean calibration instead gives factors of 0.993, 0.969, and 0.857 and capacities of 2.501, 2.248, and 1.672~\MW{}. Only 10, 9, and 11 of 15 origins individually meet the 0.95 target, and the worst coverages are 0.842, 0.753, and 0.552. The reported derating matches mean coverage, not a guarantee for every origin; shorter retraining or a more conservative rule would be needed for that guarantee.

Common flexibility representations can be placed on one boundary. Table~\ref{tab:formulations} instantiates them on the same fleet series at the marginal-PUE boundary, where one removed workload megawatt is one facility megawatt; assumed tiers act on the central facility load series (52.32~\MW{} mean, four-hour 95\% running minimum 44.11~\MW{}), trace-derived rows on eligible GPU relief. The 30/50/20 row treats its 50\% flexible tranche as shiftable within the day, an assumed envelope that this trace does not support.

\begin{table}[t]
\caption{Four-hour, 95\% relief implied by flexibility representations on the same fleet series at the marginal-PUE boundary, and their ratio to the trace-derived value at one, four, and 24 hours. Ratios are point-estimate quotients on the 15-cluster series.}
\label{tab:formulations}
\centering
\scriptsize
\begin{tabular}{@{}llcc@{}}
\toprule
Representation & Applied to & $K_{0.95,4}$ & Ratio 1/4/24~h \\
\midrule
Fully curtailable (Duke-style) & facility & 44.1 & 17.7 / 19.0 / 22.3 \\
30/50/20 flex.+interr.\ (assumed) & facility & 30.9 & 12.4 / 13.3 / 4.5 \\
20\% interruptible tier & facility & 8.8 & 3.5 / 3.8 / 4.5 \\
Attributed boundary (no retained idle) & GPU & 4.19 & 1.74 / 1.81 / 1.90 \\
Const.\ share, mean-calibrated (12.1\%) & GPU workload & 2.93 & 1.18 / 1.26 / 1.47 \\
Trace-derived, idle retained & GPU & 2.32 & 1 \\
Const.\ share, 4-h tail-calibrated (9.6\%) & GPU workload & 2.32 & 0.93 / 1.00 / 1.17 \\
\bottomrule
\end{tabular}
\end{table}

\begin{figure*}[t]
  \centering
  \includegraphics[width=0.60\textwidth]{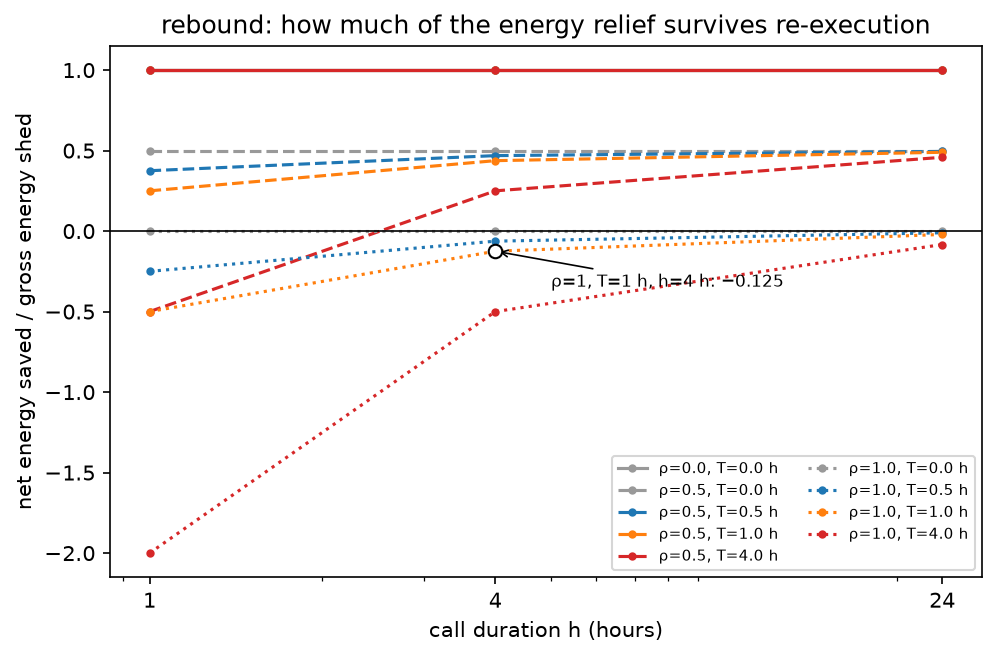}
  \caption{Net energy change across event durations, rebound fractions, and checkpoint intervals. Positive curtailed energy need not imply energy savings.}
  \Description{Lines show the fraction of gross curtailed energy that survives rebound across one-, four-, and twenty-four-hour calls, three rebound fractions, and four checkpoint intervals.}
  \label{fig:energy}
\end{figure*}

Recovery can reverse the energy interpretation of curtailment. For an $h$-hour call, rebound fraction $\rho$, and checkpoint interval $T_{\mathrm{ckpt}}$, expected lost work of $T_{\mathrm{ckpt}}/2$ gives net-to-gross energy ratio $1-\rho[1+T_{\mathrm{ckpt}}/(2h)]$. At $h=4$, $\rho=1$, and $T_{\mathrm{ckpt}}=1$ hour, the ratio is $-0.125$ (Fig.~\ref{fig:energy}). Recovery time must use the offered $qK_{0.95,4}$ rather than the larger mean eligible layer. At $q=1$, its 10.44~\MWh{} rebound takes 4.49 hours when absorbed at 2.326~\MW{}, set to 10\% of the 23.264~\MW{} mean floor-plus-shift power; the resulting 8.49-hour spacing permits at most 19.8 four-hour calls per week. Both the absorption rate and checkpoint interval are illustrative contract parameters, not measured operating limits.

\section{Illustrative MISO South Peak-Cap Screen}\label{app:interconnection}

The interconnection screen tests residual-load accounting under a simple regional cap. Hourly demand $D_t$ covers 17,544 hours of MISO South in 2024--2025, with a 32.436~GW peak and 20.202~GW mean~\cite{eia930}. For margin $m$, the cap is $P_{\mathrm{cap}}=\max_tD_t(1+m)$. A candidate load of empirical-peak scale $L$ follows normalized trace shape $s_t$ and has normalized curtailable profile $c_t$. It is admissible only if residual excess $r_t=\max(0,D_t+Ls_t-P_{\mathrm{cap}})$ satisfies $r_t\leq Lc_t$ in every hour and if the number of hours with $r_t>0$ stays within the response budget. The budget limits calls; it never permits post-response cap violations.

The 185-day data-center series is circularly shifted and tiled to the two-year demand record for each of 200 random offsets. Its scale is the seven-day rolling maximum of central facility power, an empirical peak rather than nameplate. For the trace-derived cases, we first convert eligible workload power to facility response and then divide it by the same facility empirical peak used for $s_t$. These cases set $q=1$; a field-measured $q$ would scale the capability profile before screening. The main screen adopts a conservative marginal response of 1.0 facility MW per removed workload MW; an average-PUE sensitivity uses 1.2. The remaining cases set $c_t$ to the full load, a constant 20\% tier, the mean trace-eligible share, or the attributed upper bound. We test margins of 0, 2, 5, and 10\% and annual response budgets of 0.25--2\% of hours.

\begin{figure*}[t]
  \centering
  \includegraphics[width=0.91\textwidth]{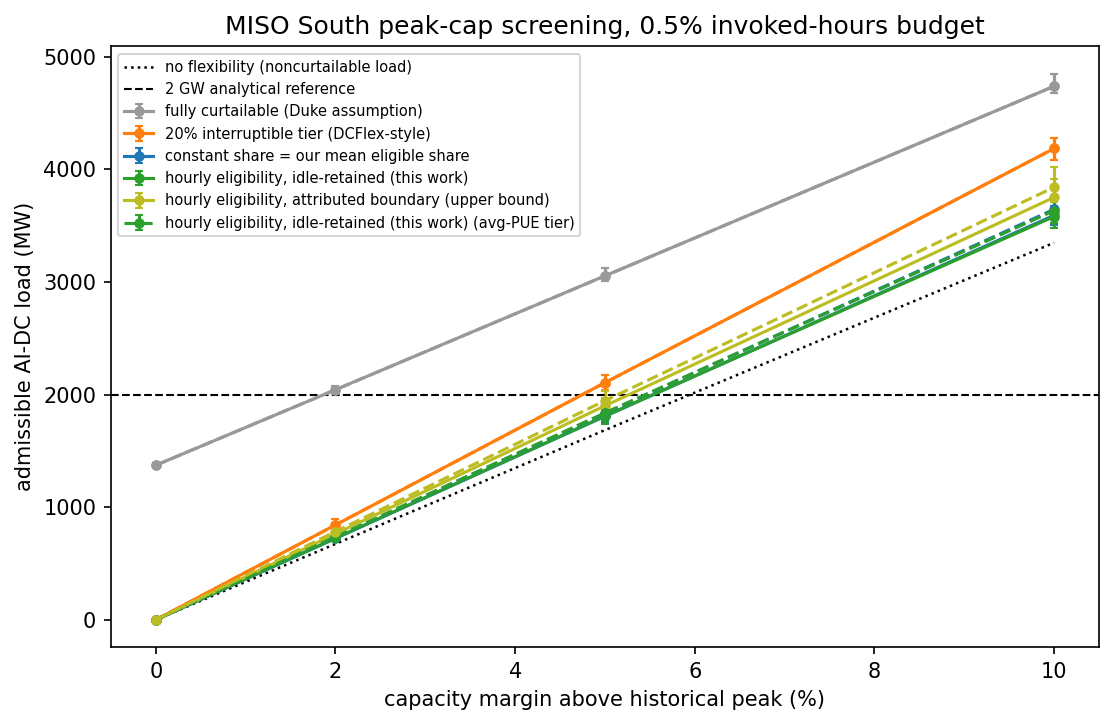}
  \caption{Illustrative peak-cap headroom in MISO South across capacity margins and response representations. Trace-derived curves use historical availability at $q=1$ and the conservative 1.0-MW marginal facility response; the 1.2 average-PUE response is a sensitivity. Values are medians over 200 trace alignments; bands show P5--P95. The calculation omits network and contingency constraints.}
  \Description{Lines show screened gigawatts increasing with regional capacity margin. Fully curtailable and fixed 20-percent assumptions lie above the trace-derived and noncurtailable-load cases.}
  \label{fig:headroom}
\end{figure*}

The 5\%-margin, 0.5\%-budget case illustrates why the residual floor matters (Fig.~\ref{fig:headroom}). A noncurtailable load screens at 1.684~GW; under the main marginal-response convention, idle-retained eligibility raises the median to 1.809~GW (1.742--1.865~GW across offsets), a 7.43\% gain. Applying the 1.2 average-PUE response instead gives 1.837~GW (1.762--1.899~GW), a 9.07\% gain. The mean-share scalar gives 1.807~GW under the main convention, effectively the same answer in this alignment-and-cap screen but not a substitute for the duration analysis. The 20\% and full-curtailment cases give 2.105 and 3.054~GW, 16.35\% and 68.76\% above the hourly trace result. At zero margin, every representation with a noncurtailable floor admits zero load, while the fully curtailable case admits 0.94--3.09~GW depending on its hour budget.

The response-hour budget is slack in the 5\%-margin trace-derived case. Evaluating each of the 200 alignments at its own admissible load produces a median of three response events (P5--P95: two--three), three invoked hours (two--four), and a longest event of two hours (one--two). The nominal 0.5\% budget permits 87.72 hours. This short-event distribution explains why the mean-share and hourly-trace screens nearly coincide in this case; it does not establish that a scalar will remain accurate for an externally specified multi-hour product.

\begin{figure*}[t]
  \centering
  \begin{minipage}[t]{0.49\textwidth}
    \centering\includegraphics[width=\linewidth]{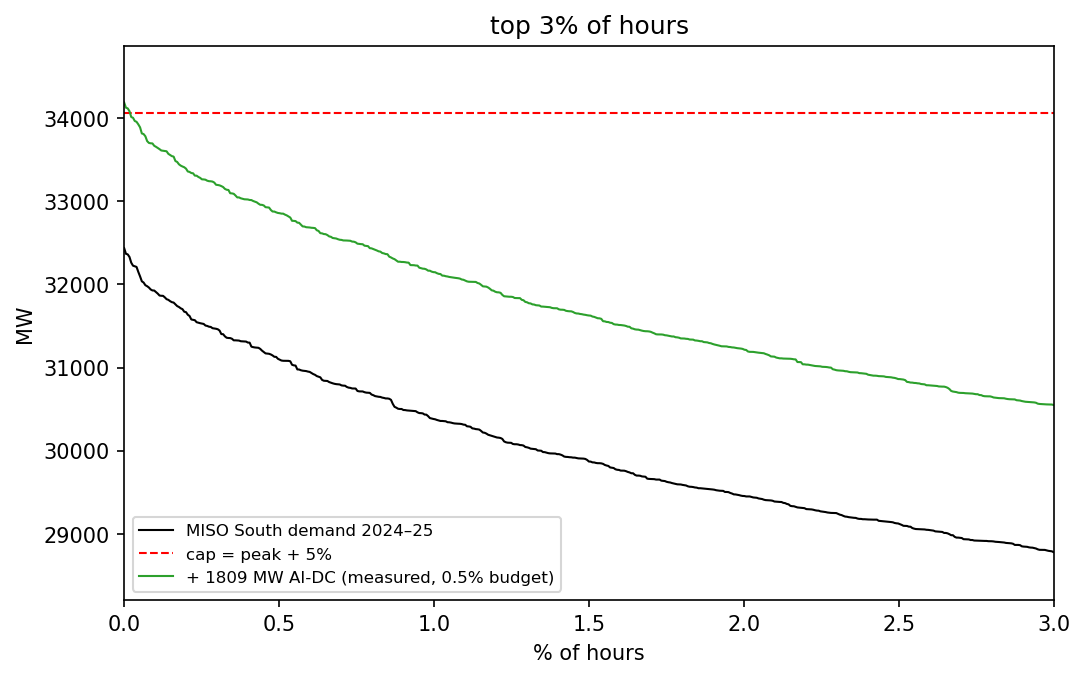}
  \end{minipage}\hfill
  \begin{minipage}[t]{0.49\textwidth}
    \centering\includegraphics[width=\linewidth]{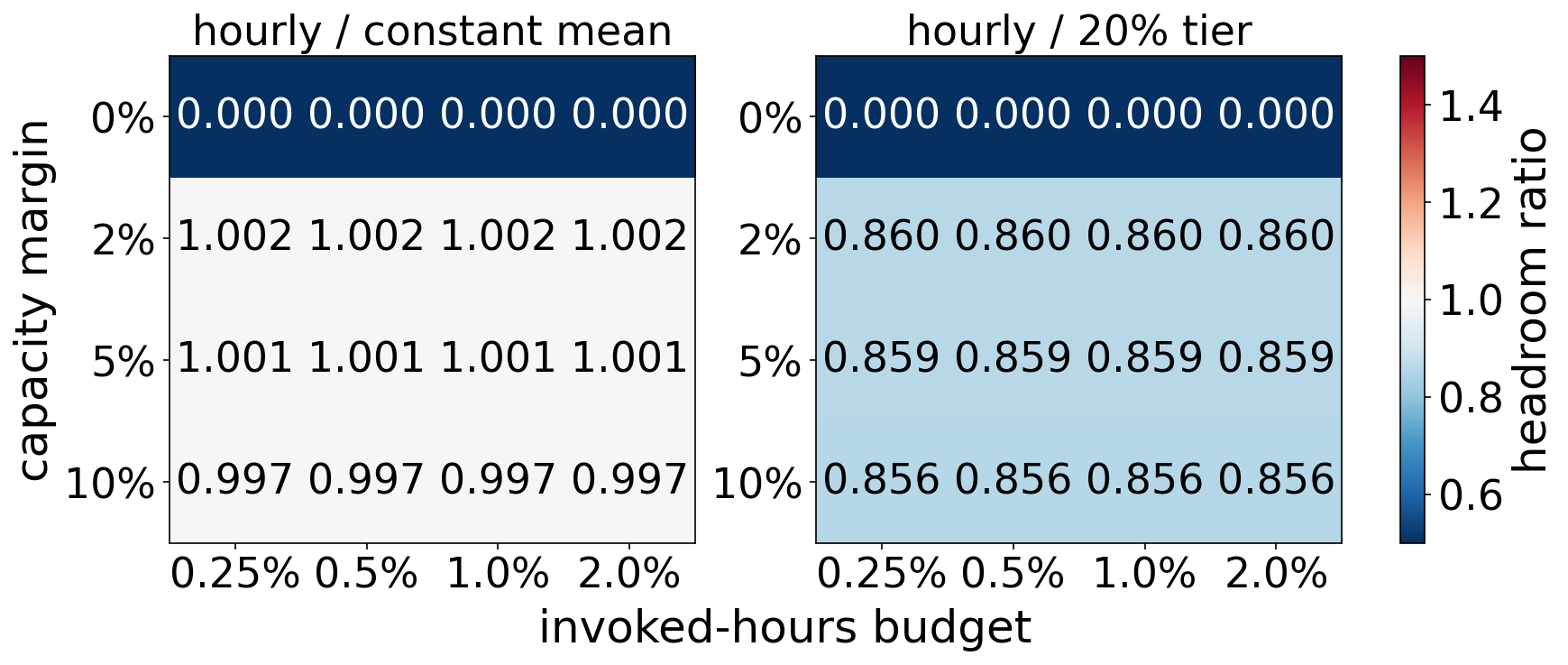}
  \end{minipage}
  \caption{Additional peak-cap results. (a) Top of the MISO South load-duration curve. (b) Headroom ratios across margins and response budgets.}
  \Description{The left panel zooms into the highest regional demand hours and the capacity cap. The right panel is a heat map of headroom ratios for flexibility representations across margins and hour budgets.}
  \label{fig:headroomextra}
\end{figure*}

The load-duration and ratio views expose which constraint binds. Figure~\ref{fig:headroomextra} focuses on the upper tail of regional load and compares representation ratios across margins and budgets. Under the trace-derived envelope, the tested annual hour budgets do not change headroom in the 5\%-margin case, so the residual floor binds before the hour allowance. The screen supports joint residual-floor and call-budget terms, not replacement of one by the other.

\end{document}